\documentclass[journal]{IEEEtran}
\ifCLASSINFOpdf
\else
\fi
\usepackage{amsmath}
\usepackage{verbatim}
\usepackage{graphicx}
\usepackage{subfigure}
\usepackage{multicol}  
\usepackage{multirow}
\usepackage{booktabs}
\usepackage{threeparttable}
\usepackage{siunitx}
\usepackage{url}
\usepackage{bm}
\usepackage{array}
\usepackage{cite}

\usepackage{color}

\begin{document}

%
\title{Modeling of Individual HRTFs based on Spatial Principal Component Analysis}
%
%
%
\author{Mengfan~Zhang,~
        Zhongshu~Ge,~
        Tiejun~Liu,~
        Xihong~Wu
        and~Tianshu~Qu~
\thanks{Copyright 2020 IEEE. Personal use of this material is permitted.  Permission from IEEE must be obtained for all other uses, in any current or future media, including reprinting/republishing this material for advertising or promotional purposes, creating new collective works, for resale or redistribution to servers or lists, or reuse of any copyrighted component of this work in other works. 

Manuscript received June 6, 2019; revised September 17, 2019 and December 15, 2019; accepted January 11, 2020. Date of publication January 17, 2020; date of current version February 4, 2020. This work was supported by the National Natural Science Foundation of China under Grant 11590773, Grant 61175043, and Grant 61421062, and in part by High-performance Computing Platform of Peking University. The associate editor coordinating the review of this manuscript and approving it for publication was Prof. A. Sarti. (Corresponding author: Tianshu Qu.)

M. Zhang, Z. Ge, X. Wu, and T. Qu are with the Key Laboratory on Machine Perception (Ministry of Education), Speech and Hearing Research Center, Peking University, Beijing 100871, China (e-mail: zhangmengfan@pku.edu.cn; gezhongshu@pku.edu.cn; wxh@cis.pku.edu.cn; qutianshu@pku.edu.cn).

T. Liu is with the State key Laboratory of Robotics, Shenyang Institute of Automation, Chinese Academy of Sciences, Shenyang, Liaoning 110004, China (e-mail: tjliu@sia.cn).

Digital Object Identifier 10.1109/TASLP.2020.2967539}}

%
%

\markboth{IEEE/ACM Transactions on Audio, Speech and Language Processing,~Vol.~28, No.~1, 2020}%
{Shell \MakeLowercase{\textit{et al.}}: Bare Demo of IEEEtran.cls for IEEE Journals}
%



\maketitle

\begin{abstract}
Head-related transfer function (HRTF) plays an important role in the construction of 3D auditory display. This paper presents an individual HRTF modeling method using deep neural networks based on spatial principal component analysis. The HRTFs are represented by a small set of spatial principal components combined with frequency and individual-dependent weights. By estimating the spatial principal components using deep neural networks and mapping the corresponding weights to a quantity of anthropometric parameters, we predict individual HRTFs in arbitrary spatial directions. The objective and subjective experiments evaluate the HRTFs generated by the proposed method, the principal component analysis (PCA) method, and the generic method. The results show that the HRTFs generated by the proposed method and PCA method perform better than the generic method. For most frequencies the spectral distortion of the proposed method is significantly smaller than the PCA method in the high frequencies but significantly larger in the low frequencies. The evaluation of the localization model shows the PCA method is better than the proposed method. The subjective localization experiments show that the PCA and the proposed methods have similar performances in most conditions. Both the objective and subjective experiments show that the proposed method can predict HRTFs in arbitrary spatial directions.
\end{abstract}


%
\IEEEpeerreviewmaketitle

\section{Introduction}
%
%
%
%
\IEEEPARstart{T}{he} head-related transfer function (HRTF) describes the acoustic transmission of sound waves from a sound source to a listener's binaural ears. It assumes the head exists in a free field. In time domain, HRTF is called head-related impulse response (HRIR) \cite{blauert1997spatial}. HRTF has been widely used in virtual sound technology, room acoustics simulation, multimedia, and virtual reality. Measuring the high spatial resolution HRTFs for each potential user is difficult, so scholars basically use non-individual HRTFs. However, using non-individual HRTFs may lead to some perception errors such as in-head localization, front-back confusion, and a breakdown of elevation discrimination ability \cite{Wenzel1991Localization,wenzel1993localization}. Thus, attaining individual HRTFs is very important and urgent in virtual auditory scene synthesis.

At present, experimental measuring is an accurate method to obtain individual HRTFs. In the past two decades, a number of research groups have performed HRTF measurements and established HRTF databases \cite{algazi2001cipic,Qu2009Distance}. However, experimental measuring of individual HRTFs requires rigorous experimental conditions and complicated equipment and keeps the subjects not moving during the measuring procedure, which make this method hard to implement.

With the development of computer technology, numerical calculation can be used to obtain HRTFs. Common numerical calculation methods include the boundary element method (BEM) \cite{Kreuzer2009Fast}, the finite element method (FEM) \cite{ma2015finite} and the finite difference method (FDM) \cite{Xiao2003Finite}. However, numerical calculation methods are computationally expensive and depend on the availability of precise 3D geometry. For example, magnetic resonance imaging (MRI) is used to obtain individual morphology. This method requires a non-trivial acquisition process and complicated calculation. 

In recent years, many researchers have concentrated on modeling individual HRTFs. Brown et al. \cite{brown1998structural} separated the effects of different physiological structures on HRTFs, and each effect was modeled with a low-order sub-filter. The combination of all sub-filters represents an HRTF. Middlebrooks \cite{Middlebrooks1999Individual} used the frequency scaling method, assuming that the HRTF spectral characteristics of diverse individuals are similar, but the corresponding frequencies of spectral characteristics are different. Through the frequency scaling method, a new subject\rq s HRTF can be obtained. Zotkin et al. \cite{Zotkin2003HRTF} selected the HRTF data of the subject whose anthropometric parameters were closest to the new subject. Jin et al. \cite{jin2000enabling} applied the principal component analysis (PCA) to the HRTF amplitude spectrum and anthropometric parameters separately and then constructed a linear mapping from the PCA weights of the anthropometric parameters to the PCA weights of HRTFs. Hu et al. \cite{Hu2008HRTF} used back-propagation artificial neural networks to map the PCA weights of HRTFs to the selected anthropometric parameters. However, this approach required separately training neural networks for each spatial direction. To predict high spatial resolution HRTFs, this method had to measure a large HRTF database and train thousands of neural network models. 

As the subjective perception of spatialization is the ultimate goal, the individual HRTFs can also be obtained based on the listener’s feedback \cite{guezenoc2018hrtf}. Fink et al. \cite{Fink2012Tuning} let subjects tune the PCA weights from average HRTFs to obtain individual HRTFs. This tuning procedure can reduce localization errors; however, obtaining a customized HRTF for a subject is very time-consuming. The subjects need a lot of time to finish the tuning part. Luo et al. \cite{luo2013virtual} also used the tuning method to obtain individual HRTFs and first introduced deep learning autoencoders to HRTF. The autoecoder was used to perform feature reduction and obtained a better result than PCA.

The aim of this paper is to realize the modeling of individual HRTFs and to predict the HRTFs in arbitrary spatial directions. Spatial principal component analysis (SPCA) \cite{Xie2012Recovery} or spherical harmonics (SHs) basis functions \cite{romigh2015efficient} can be used to spatially decompose HRTFs. In this study, we use SPCA to decompose HRTF into a weighted combination of spatial principal components (SPCs). Through the deep neural network (DNN) training, the SPCs in arbitrary spatial directions are estimated. A small quantity of anthropometric parameters were selected and mapped to the SPCA weights using neural networks. Then, we combined the predicted SPCs and the SPCA weights for each individual to reconstruct the HRTFs in arbitrary spatial directions. 

The rest of the paper is organized as follows. In Section \ref{sec:spca}, the data preprocessing and SPCA are performed. In Section \ref{sec:model}, modeling individual HRTFs based on SPCA using DNN is described. In Section \ref{sec:obj}, the objective experiments are conducted and the objective error is analyzed. In Section \ref{sec:sub}, the subjective evaluation of the proposed approach is described. In Section \ref{sec:conclusion}, the conclusion is presented.

\section{SPCA and Data Preprocessing}
\label{sec:spca}

\subsection{CIPIC Database}
\label{sec:cipic}

The HRIRs used in this paper are derived from the CIPIC database \cite{algazi2001cipic}, which is measured by U. C. Davis CIPIC Interface Laboratory. In this database, a blocked ear technique is performed for 45 subjects (27 males, 16 females, 2 KEMARs), and 1250 directions of HRTF data were measured for each subject. Sound source directions are in interaural-polar coordinate. The database also contains up to 27 anthropometric parameters for each subject.


\subsection{Data Preprocessing}
\label{sec:hrtf_pre}

We first transform the raw HRIRs in CIPIC database into the frequency domain. Fourier transformation is applied to the HRIRs to calculate the HRTFs. Then we transform the HRTFs into a logarithmic scale. Because a logarithmic scale is much closer to our auditory perception \cite{smith1983techniques}. Therefore, the base 10 log-magnitude responses of HRTFs are computed.
\begin{equation}
HRTF_\text{log}(\theta,\varphi,f,s)=20log_{10}(|HRTF(\theta,\varphi,f,s)|).
\end{equation}

Then the mean logarithmic HRTFs is calculated from all the $HRTF_{log}$. The mean function includes the direction and subject independent spectral features shared by all HRTFs in the CIPIC database. 

\begin{equation}
\mu(f)=\frac{1}{S\times D}\sum_{s}\sum_{\theta}\sum_{\varphi}HRTF_\text{log}(\theta,\varphi,f,s),
\end{equation}

After removing the mean logarithmic HRTFs, we obtain the log-magnitude function, which is called $HRTF_{\text{log}{\Delta}}$.

\begin{equation}
HRTF_{\text{log}{\Delta}}(\theta,\varphi,f,s)=HRTF_\text{log}(\theta,\varphi,f,s)-\mu(f).
\end{equation}

\subsection{SPCA}
\label{sec:sub_spca}

PCA is mathematically defined as an orthogonal linear transformation that transforms the data to a new coordinate system, such that the greatest variance by some projection of the data comes to lie on the first coordinate (called the first principal component), the second greatest variance on the second coordinate, and so forth \cite{jolliffe2003principal}. 

The traditional PCA method is generally used in the time or frequency domain of HRTFs \cite{Kistler1992A,Zhang2011Statistical}. In contrast to traditional PCA models, SPCA is applied to the spatial domain. The high spatial resolution HRTFs can be represented as the weighted combination of orthonormal SPCs \cite{Xie2012Recovery}. The SPCA applied to a $HRTF_{\text{log}{\Delta}}$ is shown below.
\begin{equation}
\label{equ:spca}
HRTF_{\text{log}{\Delta}}(\theta,\varphi,f,s)=\sum_{q=1}^D d_q(f,s)W_q(\theta,\varphi)+H_\text{av}(\theta,\varphi),
\end{equation}
where $q$ is the identification of the SPC, $W_q$ is the SPC that depends only on the source direction, $\varphi$ is the elevation angle, and $\theta$ is the azimuth angle. $D$ is the number of spatial directions. $d_q(f,s)$ is the SPCA weight which varies as function of frequency $f$ and individual $s$. $H_\text{av}$ is the mean $HRTF_{\text{log}{\Delta}}$ magnitude across the frequencies and subjects and can be calculated as follows:
\begin{equation}
H_\text{av}(\theta,\varphi)=\frac{1}{N\times S}\sum_{s} \sum_{f} HRTF_{\text{log}{\Delta}}(\theta,\varphi,f,s),
\end{equation}
where $N$ and $S$ are the total number of frequencies and subjects respectively.

To calculate the SPCs and SPCA weights, we combine $HRTF_{\text{log}{\Delta}}$ of all the frequencies, directions and subjects into an $(NS)\times{D}$ matrix $\bm{H}$. Each column of $\bm{H}$ corresponds to a spatial direction, and each row of $\bm{H}$ represents the HRTF of an individual at a discrete frequency. We subtract the mean value from $\bm{H}$ to obtain $\bm{H_\Delta}$. Then we calculate the covariance matrix $\bm{R}$:
\begin{equation}
\bm{R}=\bm{H_\Delta^T}\bm{H_\Delta},
\end{equation}
where $\bm{R}$ is a $D\times{D}$ matrix. Its eigenvectors are extracted and arranged as the eigenvalue reduced-order. Then the first $Q$ eigenvectors are taken as the base vectors, i.e. the SPCs which represent the values of basis functions at $D$ discrete directions, and form a $Q\times{D}$ matrix $\bm{W}$:
\begin{equation}
\bm{W}=[W_q(0),W_q(1),...,W_q(D-1)].
\end{equation}
Each row of $\bm{W}$ corresponds to a SPC, and each column of $\bm{W}$ represents the values of SPCs at a spatial direction. We name the column of $\bm{W}$ as direction vector of SPCs (DV-SPCs). Since all the SPCs are orthogonal to one another, by the use of these SPCs, we can obtain the SPCA weights:
\begin{equation}
\bm{d}=\bm{H_{\Delta}}\bm{W^T},
\end{equation}
where $\bm{d}$ is a $(NS)\times{Q}$ matrix composed of the SPCA weights for all the individuals and frequencies, $\bm{W^T}$ is the transpose of $\bm{W}$. Finally, we can predict all the HRTFs as
\begin{equation}
\label{equ:reconstr}
\bm{H}=\bm{d}\bm{W}+\bm{H_\text{AV}},
\end{equation}
where $\bm{H_\text{AV}}$ is an $(NS)\times{D}$ matrix. Each row of $\bm{H_\text{AV}}$ corresponds to the $H_\text{av}$ in $D$ directions, and all of the rows are identical. 

If the number of the SPCs is chosen to be equal to the number of spatial directions, i.e. $Q = D$, then the HRTFs can be fully represented without loss, as shown in Eq.(\ref{equ:spca}). If $Q < D$ is selected, then only an approximate representation of the original HRTF magnitude can be obtained. The principal component variances are the eigenvalues of the covariance matrix $\bm{R}$. The cumulative percentage of variance in reconstructed HRTF is calculated as
\begin{equation}
Var=\frac{\sum_{q=1}^Q \lambda_q}{\sum_{q=1}^D \lambda_q}\times{100\%},
\end{equation}
where $\lambda_q$ is an eigenvalue, and $Var$ is the cumulative percentage of variance. $Var$ is increased with the selection of $Q$, as shown in Table \ref{table:spca}. We can restore more than 70\% of the total variability by selecting the first 20 SPCs. We can recover more than 80\% of the total variability by selecting the first 60 SPCs. Prior studies reported that more than 90\% of the total variability is enough to recover the HRTF magnitudes when applying PCA in frequency domain \cite{Hu2008HRTF,Zeng2010A}. Therefore, we select the first 200 SPCs to recover more than 90\% of the total variability when applying PCA in spatial domain. 

\begin{table}[!t]
\centering
\caption{Cumulative percentage of variance in reconstructed HRTF magnitudes is increased with the number of selected SPCs}
\label{table:spca}
\begin{tabular}{c|cc}
\hline
\multirow{2}{*}{\begin{tabular}[c]{@{}c@{}}The number of SPCs ($Q$)\end{tabular}} & \multicolumn{2}{c}{\begin{tabular}[c]{@{}c@{}}Cumulative percentage\\ of variance (\%)\end{tabular}} \\ \cline{2-3} 
                                                                                                     & \multicolumn{1\quad}{c|}{Left ear}                                    & Right ear                                   \\ \hline
1                                                                                                    & \multicolumn{1}{c|}{16.54}                                       & 20.14                                       \\
5                                                                                                    & \multicolumn{1}{c|}{52.20}                                        & 55.33                                       \\
10                                                                                                   & \multicolumn{1}{c|}{62.29}                                       & 64.84                                       \\
20                                                                                                   & \multicolumn{1}{c|}{70.10}                                        & 71.85                                       \\
50                                                                                                   & \multicolumn{1}{c|}{78.33}                                       & 79.54                                       \\
60                                                                                                   & \multicolumn{1}{c|}{80.09}                                       & 81.22                                       \\
80                                                                                                   & \multicolumn{1}{c|}{82.93}                                       & 83.98                                       \\
100                                                                                                  & \multicolumn{1}{c|}{85.11}                                       & 86.09                                       \\
200                                                                                                  & \multicolumn{1}{c|}{91.03}                                       & 91.56                                       \\
500                                                                                                  & \multicolumn{1}{c|}{97.07}                                       & 97.22                                       \\ \hline
\end{tabular}
\end{table}

\subsection{Key Anthropometric Parameters Selection}
\label{sec:hrtf_data_pre}

There are 27 anthropometric parameters in the CIPIC database. Measuring all the anthropometric parameters for each individual is a time-consuming and difficult process. In addition, not all of these anthropometric parameters are strongly related to HRTFs. Therefore, it is necessary to select a small set of anthropometric parameters which are most strongly related to the variations in the HRTFs of different individuals \cite{Hu2008HRTF,Zhang2011Statistical}. We process the HRIRs and the anthropometric parameters in CIPIC database as follows.

First, Fourier transformation is applied to the HRIRs in the CIPIC database, and the mean of the obtained HRTF data is subtracted. For each sampled direction in the CIPIC database, we apply the traditional PCA to the HRTF data of all the subjects,
\begin{equation}
\label{equ:pca}
HRTF(f,s)=\sum_{q=1} ^N d_q(s)W_q(f)+H_\text{av}(f),
\end{equation}
where $W_q$ is the PC. Note that the traditional PCA needs to model HRTFs in each measured spatial direction, which means we apply PCA 1250 times in total. The resultant PCs and $H_\text{av}$ depend only on the frequency, whereas the SPCs and $H_\text{av}$ obtained by SPCA depend only on spatial directions. $d_q(s)$ is the PCA weight which varies as function of individual $s$.

Second, multiple linear regression analysis is used to analyze the relationship between the PCA weights and the anthropometric parameters. 
\begin{equation}
d_q=s\gamma+e,
\end{equation}where $\gamma$ is the regression coefficient, and $e$ is the error. Then we use $t$ statistics to identify which parameters have a significant effect on the PCA weights.

Third, Pearson correlation coefficient analysis is introduced to measure the strength of dependence between each of the two anthropometric parameters,
\begin{equation}
r_{ij}=\left|\frac{\sum(s_i-\overline{s_i})(s_j-\overline{s_j})}{\sqrt{(\sum(s_i-\overline{s_i})^2(s_j-\overline{s_j})^2)}}\right|,
\end{equation}
where $s$ is a vector that comprises 27 anthropometric parameters, $i,j=1,2,...,27$ and $i\neq j$. $r_{ij}$ is the degree of correlation between two anthropometric parameters $s_i$ and $s_j$. This procedure is used to reduce the number of parameters to make the measurement more feasible. A stronger correlation indicates that one parameter could be represented by another \cite{Zeng2010A}.

\begin{table}[!t]
\centering
\caption{Selected key anthropometric parameters}
\label{table:parameter}
\begin{tabular}{c|c|c|c}
\hline
Variable & Measurement      & Variable  & Measurement           \\ \hline
$x_1$      & head width        & $d_1$        & cavum concha height   \\
$x_2$*     & head height       & $d_3$        & cavum concha width    \\
$x_3$      & head depth        & $d_4$        & fossa height          \\
$x_{12}$     & shoulder width    & $d_5$        & pinna height          \\ 
        &                   & $d_6$        & pinna width           \\ \hline
\end{tabular}\\
\vspace{2pt}
\centerline{* This parameter is only used in the prediction of ITDs.}
\end{table}

\begin{figure}[!t]
 \center{\includegraphics[width=5cm] {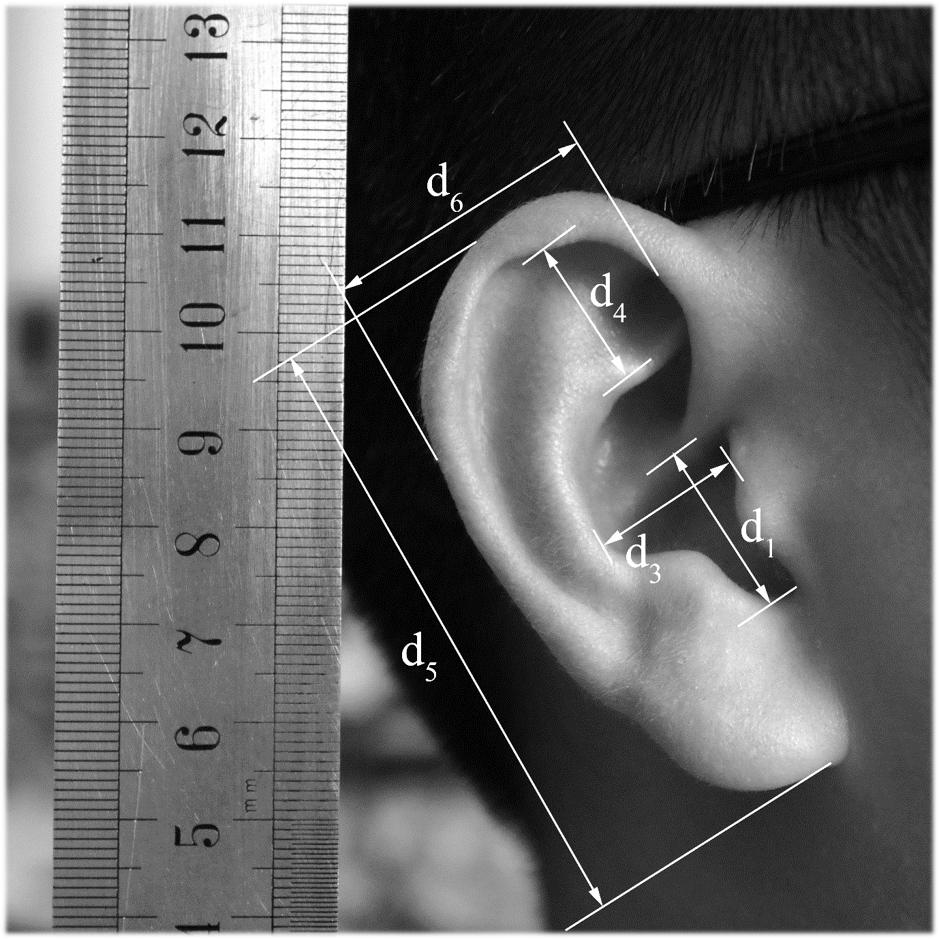}}
 \caption{\label{fig:ear} Five selected pinna parameters in our model.}  
\end{figure}

Finally, based on the correlation between the anthropometric parameters as well as the analysis of the PCA weights and anthropometric parameters, we balance the principle of simpleness, completeness and feasibility in practice to reduce parameters. Anthropometric parameters with large correlation are reduced while considering the theoretical and practical influence of them on HRTFs to determine which to be remained.  Eight anthropometric parameters that have a significant effect on the PCA weights are finally selected. The quantities of selected anthropometric parameters are the same as \cite{Hu2008HRTF,Zhang2011Statistical}. As shown in Table \ref{table:parameter}, the eight anthropometric parameters are head width, head depth, shoulder width, cavum concha height, cavum concha width, fossa height, pinna height and pinna width, corresponding to $x_1$,  $x_3$,  $x_{12}$,  $d_1$, $d_3$, $d_4$, $d_5$ and $d_6$ in the CIPIC database. Note that the head height parameter $x_2$ is only used in the prediction of ITDs. The first four anthropometric parameters are head and torso parameters, which can be measured by a caliber rule or taking pictures. The last five anthropometric parameters are pinna parameters, which can be obtained by photograph annotation as shown in Fig. \ref{fig:ear}.

\section{Modeling of Individual HRTFs}
\label{sec:model}

\subsection{Outline}
\label{sec:outline}

In this paper, the magnitude spectra of HRTFs are modeled based on SPCA using neural networks. The phase of HRTFs is calculated by minimum-phase reconstruction \cite{Kistler1992A}. 

\begin{figure}[!t]
 \center{\includegraphics[width=8cm] {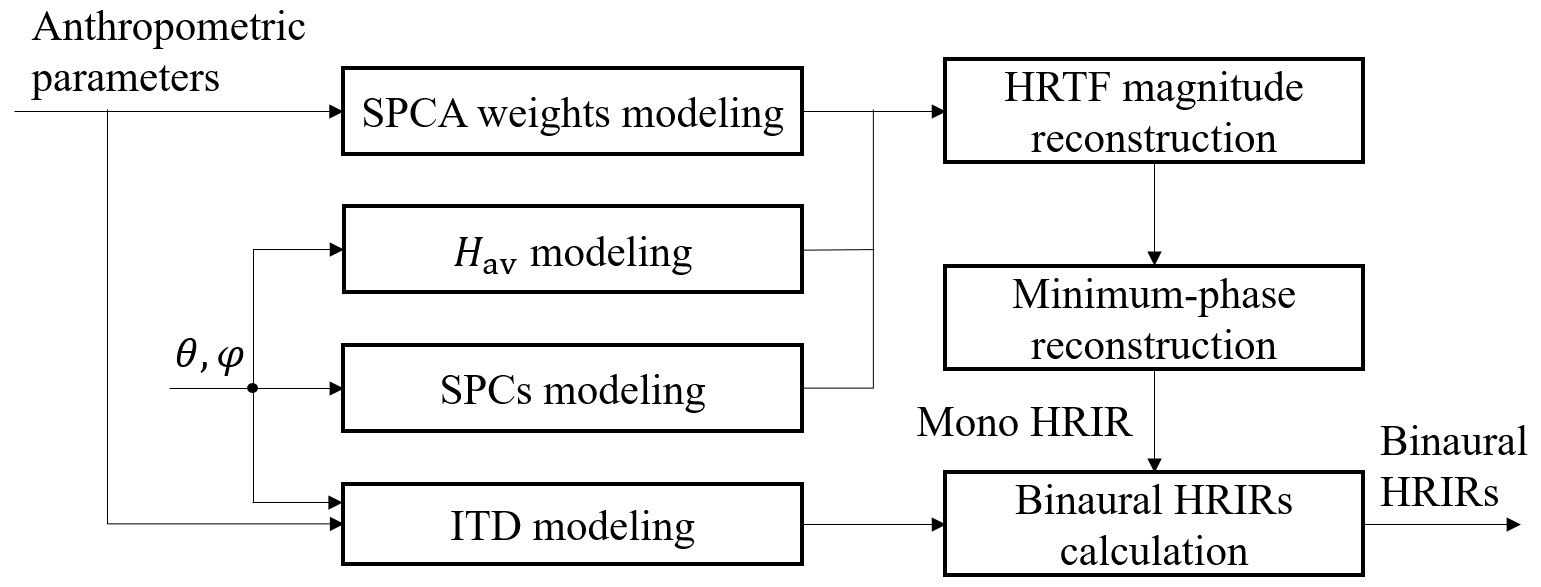}}
 \caption{\label{fig:frame} The framework of individual HRTF modeling.}      
\end{figure}

Fig. \ref{fig:frame} depicts the framework of individual HRTF modeling. First we model the SPCA weights, the SPCs, the $H_\text{av}$ and the ITDs, respectively. The SPCA weights can be obtained by modeling the individual's morphology, for the SPCA weights vary as functions of anthropometric parameters and frequency. A quantity of key anthropometric parameters is selected in Section \ref{sec:hrtf_data_pre} to represent the human morphology, therefore, the SPCA weights for any individual outside the database can be estimated from a small set of anthropometric measurements. DNNs are used to predict the SPCs, the $H_\text{av}$ and the ITDs in arbitrary spatial directions. Accordingly, HRTFs with high directional resolution can be recovered by solving the Eq. (\ref{equ:spca}) using the predicted SPCs, SPCA weights and $H_\text{av}$. Then the minimum-phase reconstruction method is used to generate mono HRIRs \cite{Kistler1992A}. Finally, binaural HRIRs are obtained using estimated ITDs and the corresponding left and right mono HRIRs.

Due to the symmetry between front and rear HRTFs, we model the front and rear HRTFs separately. Because HRTF dissimilarity increases with their angular difference, however, the low dissimilarity may occur for large angular difference. For instance, if the two HRTFs taken from two locations are symmetric with respect to the interaural axis, the two HRTFs will be very similar despite the large angular difference. Thus, it is preferred to consider the front and rear HRTFs separately in order to keep a confident link between the HRTF dissimilarity and the angular difference \cite{Nicol2006Looking}. Additionally, there exists some differences between an individual's two ears; we therefore consider one individual as containing two observations which should be modeled separately to obtain different SPCA weights. When modeling the SPCs, the $H_\text{av}$ and the ITDs, which are related to spatial directions, we split the data of these parameters into two parts, the front and rear parts. In the CIPIC database, the azimuth angles composed a vector $azi=[-80, -65, -55, -45 : 5 : 45,\ 55,\ 65,\ 80]$, and the elevation angles make up a vector $elev = -45+5.625\times [0:49]$. The front part contains $elev\le90$ in all of the azimuth angles, and the remaining angles belong to the rear part. Thus, each part contains a total of 1250 sets of data, including 625 sets of data for two ears. For the three parameters, the SPCs, the $H_\text{av}$ and the ITDs, we train two DNNs for each parameter in order to obtain a desirable prediction result.

To sum up, through securing a quantity of anthropometric parameters, we can reconstruct the individual's binaural HRIRs in arbitrary spatial directions. 

\subsection{Prediction of SPCA Weights for an Individual}
\label{sec:spca-weights}

Given that the SPCA weights are a function of frequency and an individual's morphology, we model the SPCA weights and eight key anthropometric parameters based on neural networks. After securing eight anthropometric parameters of an individual, we can estimate the SPCA weights for the individual in all of the frequency points using neural network models.

A total of 37 subjects $s_m(m=1,2,...,37)$ in the CIPIC database contain all of the eight anthropometric parameters. As aforementioned in Section \ref{sec:outline}, one individual's two ears are modeled separately to obtain different SPCA weights. Thus, we have 74 sets of anthropometric parameters; considering the total number of frequency points is $N=200$, we have 14800 sets of the SPCA weights. 

\begin{figure}[!t]
 \center{\includegraphics[width=5cm] {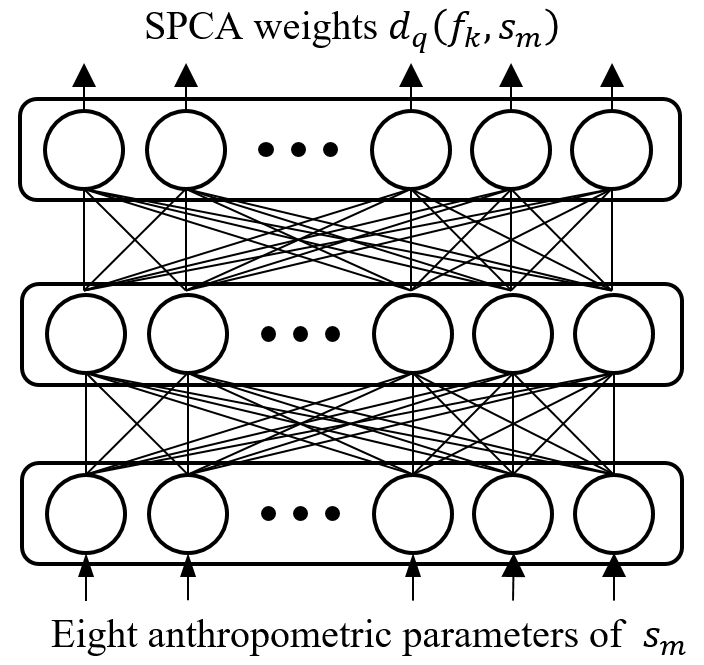}}
 \caption{\label{fig:w-neural} The architecture of neural network model for SPCA weights prediction.}      
\end{figure}

For each frequency point $f_k(k=1,2,...,N)$, we build a model for the eight anthropometric parameters and the corresponding SPCA weights. The specific architecture of the neural network model is shown in Fig. \ref{fig:w-neural}. The inputs are the eight anthropometric parameters, and the ground truth are SPCA weights in one frequency point. So in total we build 101 neural network models because of the symmetry property of the Fourier transformation. 

30 subjects' binaural data are carefully selected to guarantee that the data sets for each anthropometric parameter are widely distributed. That is to say, most of the test data are within the corresponding training data ranges \cite{Hu2008HRTF}. These data, which comprise a total of 60 sets, are used in the training phase, and 10 of them are used as the validation set to prevent over-fitting. The remaining 14 sets of data are used to test the average error of the neural network model. The mean and variance of the test set and validation set are normalized using the training set statistics to have zero mean and unit variance. All the neural network models comprise a single hidden layer, a hyperbolic tangent activation function and a hyperbolic tangent output function, and they are feedforward backpropagation neural networks with a learning rate of 0.001.

For network models of all frequencies, each network takes about 500 epochs to converge averagely. After training all the neural networks, we obtain a system for predicting the SPCA weights. The mean square error (MSE) is used to test the prediction system:
\begin{equation}
e_d=\frac{1}{Q\times{N}\times{S}}\sum\limits_{q}\sum\limits_{k}\sum\limits_{m}(\hat{d_q}(f_k,s_m)-d_q(f_k,s_m))^2,
\end{equation}
where $\hat{d_q}$ is estimated by neural network model, and $e_d$ is the reconstruction error of the SPCA weights. The MSE of the overall prediction, including all the SPCA weights for all the individuals and frequencies, is 9.25.


\begin{figure}[!t]
\centering
\includegraphics[width=9cm]{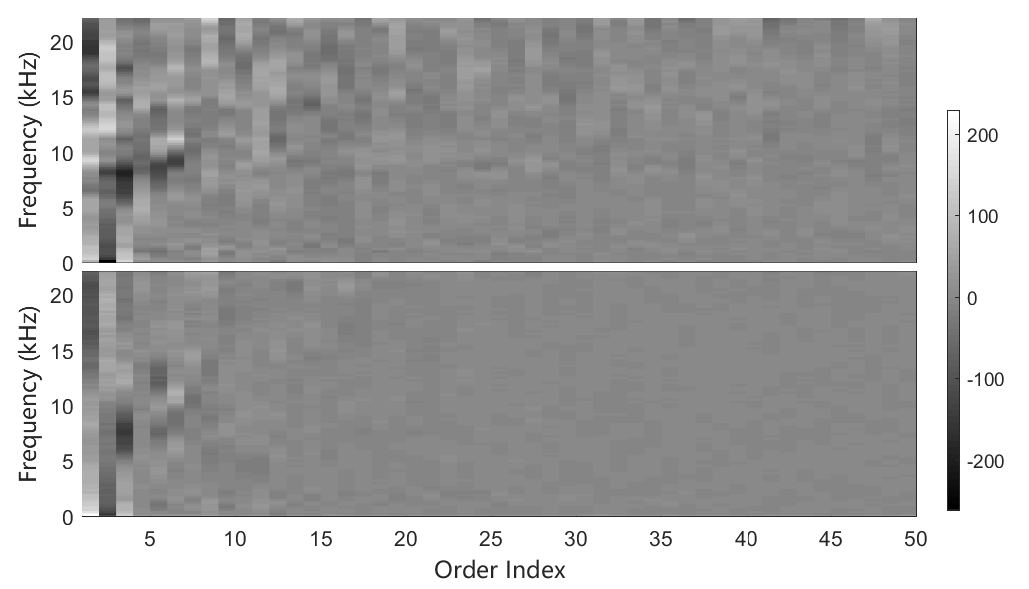}
\caption{\label{fig:weight50} Comparison of the real value (top) and the estimated value (bottom) for the first 50 orders of SPCA weights.} 
\end{figure}

\begin{figure}[!t]
\centering
\includegraphics[width=8cm]{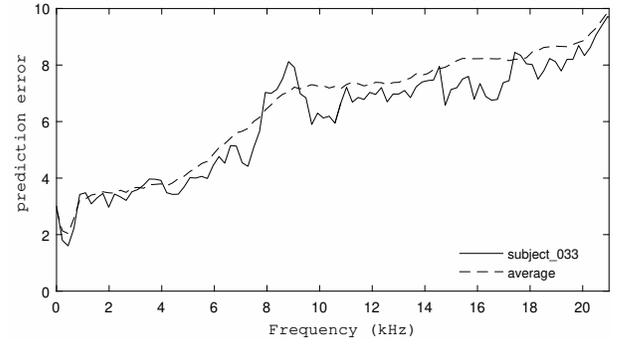}
\caption{\label{fig:weightE} The prediction errors of SPCA weights as a function of frequency.} 
\end{figure}

We randomly select a subject 033 in the CIPIC database and plot the comparison of the estimated value and real value of its left ear's first 50 orders of SPCA weights, as shown in Fig. \ref{fig:weight50}. As previously discussed in Section \ref{sec:spca}, the first 50 SPCs can restore about 80\% of the total variability, and the first 200 SPCs can reconstruct above 90\% of the total variability. Therefore, when the range of the $q$ is from 51 to 200, the total amount of variability is around 10\%. Moreover, we observe that the SPCA weights are approximately equal to 0 when $q$ is larger than 50. Thus, we only plot the curves where $q$ is less than 50 to make comparisons between the estimated SPCA weights and the real ones. Fig. \ref{fig:weightE} shows the prediction error, which is the mean of the absolute errors across all directions, of SPCA weights for subject 033 and the average prediction error. The figure shows that the difference between the real value and the estimated value of the SPCA weights increases when frequency increases. The SPCA weights in high frequency are more unstable than the lower ones. 

\subsection{Modeling of SPCs}
\label{sec:spc-model}
As discussed in Section \ref{sec:spca}, $\bm{W}$ is a $Q\times{D}$ matrix composed of the first $Q$ SPCs. Each row of $\bm{W}$ corresponds to an SPC, and each column of $\bm{W}$ represents the values of SPCs at a spatial direction. The DV-SPCs are a function of spatial directions and represented by the column of $\bm{W}$. We model the first $Q$ SPCs by predicting DV-SPCs in all sampled $D$ spatial directions, then recombining $\bm{W}$.
\begin{equation}
\bm{W} = \left[
\begin{array}{ccc}
W_1(0),\quad W_1(1) & \dots & W_1(D-1) \\
W_2(0),\quad W_2(1) & \dots & W_2(D-1) \\
\vdots \qquad\quad \vdots & \vdots & \vdots \\
W_Q(0),\quad W_Q(1) & \dots & W_Q(D-1)
\end{array} \right].
\end{equation}

\begin{figure}[!t]
 \center{\includegraphics[width=5cm] {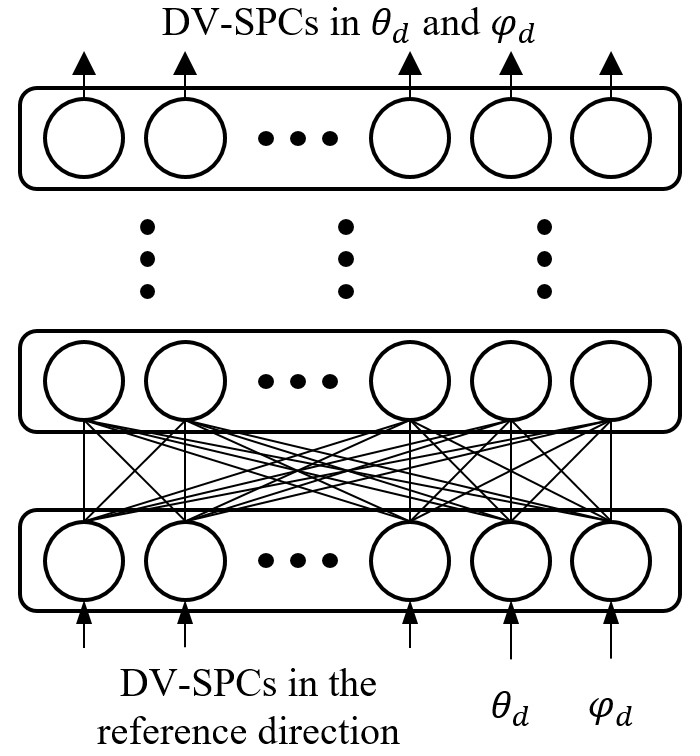}}
 \caption{\label{fig:dv-neural} The architecture of DNN for predicting the DV-SPCs.}      
\end{figure}

The DV-SPCs are modeled with DNNs. We set the angles of directly ahead ($\varphi=\ang{0}$, $\theta=\ang{0}$) and the directly behind ($\varphi=\ang{180}$, $\theta=\ang{0}$) as the reference direction for the front and rear data sets, respectively. The inputs of DNN are the DV-SPCs in the reference direction, the target azimuth $\theta_d$ and the target elevation $\varphi_d$ in degrees. We set the ground truth as the DV-SPCs in the target direction. The specific architecture of the DNN for predicting the DV-SPCs is shown in Fig. \ref{fig:dv-neural}. Inputing the DV-SPCs in the reference direction reduces the complexity of learning process. The main task of the DNN becomes learning the difference between the DV-SPCs in the reference direction and target direction. This can effectively improve the task result.

To guarantee the variability of the test data, we index the 1250 sets of data and select the test data every four indexes. The extra directions of data are then re-indexed and we choose the validation data every five indexes. The remaining directions of data are used as training data. Thus, we uniformly distribute the training set, validation set and test set in the space. 

\begin{figure}[!t]
\centering
\includegraphics[width=9cm]{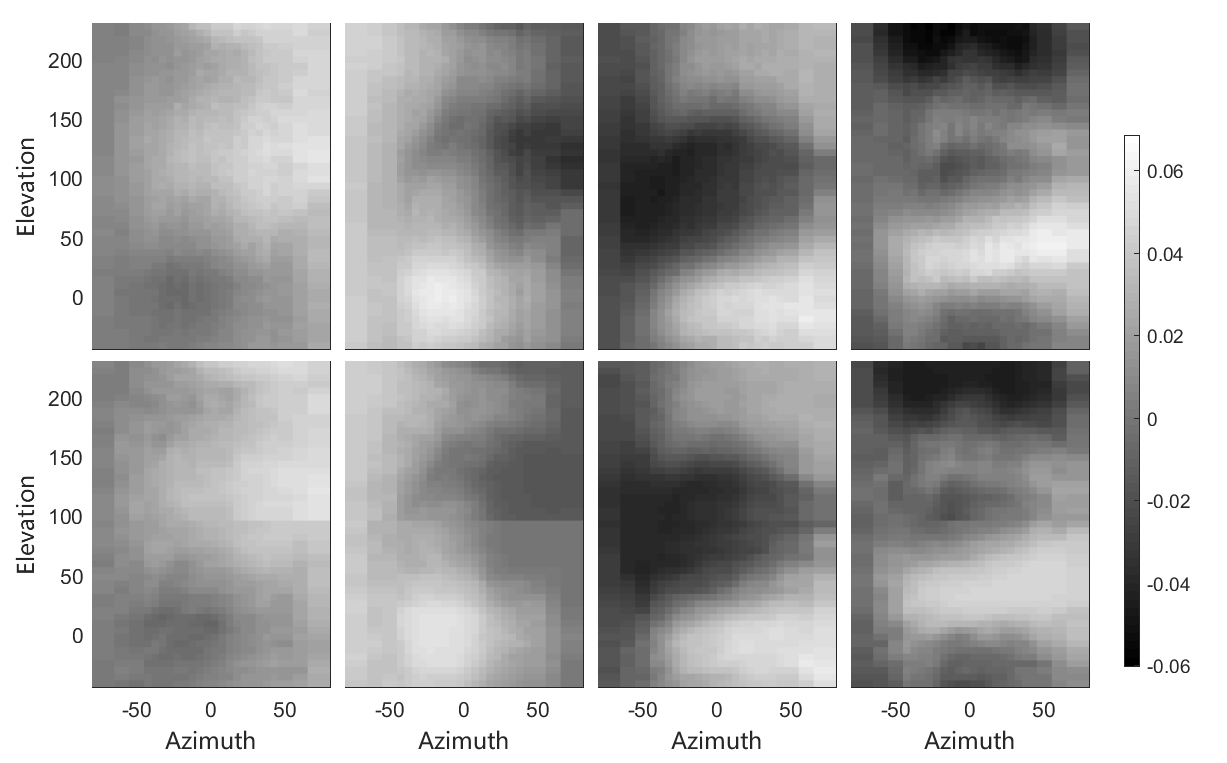}
\caption{\label{fig:6spc} Comparison of the real value (top row) and the estimated value (bottom row) of the first 4 SPCs (4 columns).} 
\end{figure}

The validation set is used to control the training phase and prevent over-fitting, and the test set is used to estimate the generalization error of the modeling. The mean and variance of the test set and validation set were normalized using the training set statistics to have zero mean and unit variance. The DNNs were implemented by MATLAB 's DeepLearnToolbox-master \cite{IMM2012-06284}. Each DNN is fully connected and has three hidden layers. Three hidden layers are chosen because it has a better performance than using only one or two hidden layers; if more than three hidden layers are used, the performance does not have a greater improvement, and the neural network is also easy to overfit.  Both the activation function and the output function are set hyperbolic tangents with the learning rate set as 0.001. These settings which were determined through the results of many experiments can lead to good prediction performance. 

It takes about 5000 epochs for the DNN models to converge. By training the DNN models, the DV-SPCs in arbitrary spatial directions can be predicted. We combine the DV-SPCs in all the $D$ sampled directions to obtain the $Q\times{D}$ matrix $\bm{W}$. MSE is used to calculate the reconstruction error:
\begin{equation}
e_W=\frac{1}{D}\sum\limits_{\theta}\sum\limits_{\phi}(\hat{W_q}(\theta,\phi)-W_q(\theta,\phi))^2,
\end{equation}
where $\hat{W_q}$ is estimated by the DNN model and $e_W$ is the reconstruction error whose value is $1.97\times{10^{-2}}$.

Fig. \ref{fig:6spc} illustrates the comparison of the first, the second, the third and the fourth SPCs' real values and estimated values. The predicted SPCs are close to the real SPCs. As discussed in Section \ref{sec:spca}, the first 5 SPCs can restore more than 50\% of the total variability, and our algorithm performs well. With an increment of $q$, the corresponding SPC gradually becomes unstable, and the difference between the predicted SPC and the real SPC widens.

\subsection{Modeling of $H_\text{av}$}
\label{sec:msf-model}

The modeling of the $H_\text{av}$ is similar to the prediction of the DV-SPCs. This model is also based on DNN. The input of DNN is the $H_\text{av}$ in the reference direction, the target azimuth and the target elevation in degrees. The ground truth is the $H_\text{av}$ in the target direction. The reference directions in addition to the selection of training set, validation set and test set are the same as in Section \ref{sec:spc-model}. The mean and variance of the test set and validation set are normalized using the training set statistics to have zero mean and unit variance. Both of the modeled DNNs are fully connected and have three hidden layers. The activation function and the output function are set hyperbolic tangent, and the learning rate is 0.001. The architecture and the parameters are determined through the results of many experiments.

\begin{figure}[!t]
 \center{\includegraphics[width=7cm] {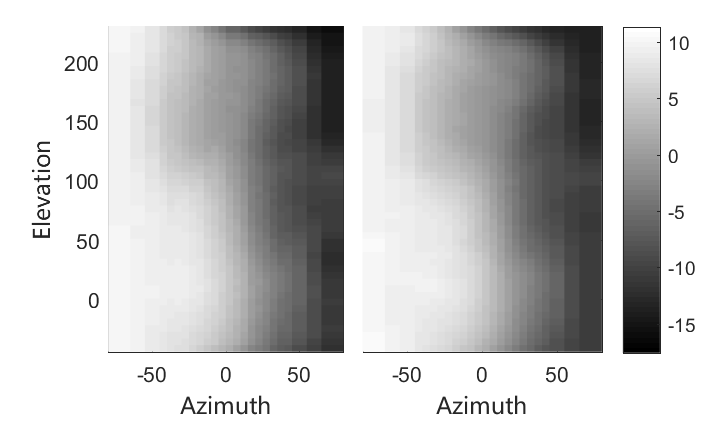}}
 \caption{\label{fig:ctf} Comparison of the real value (left) and the estimated value (right) of the $H_\text{av}$.}      
\end{figure}

It takes about 700 epochs for the DNN models to converge. After training the DNNs, we can obtain a system to predict the $H_\text{av}$ in arbitrary spatial directions. All the $D$ directions in the CIPIC database are used as the target directions to predict the corresponding $H_\text{av}$. The predicted results are shown in Fig. \ref{fig:ctf}.

The MSE is used to calculate the reconstruction error of the $H_\text{av}$:

\begin{equation}
e_H=\frac{1}{D}\sum\limits_{\theta}\sum\limits_{\phi}(\hat{H}_\text{av}(\theta,\phi)-H_\text{av}(\theta,\phi))^2,
\end{equation}
where $\hat{H}_\text{av}$ is estimated by the DNN model and $e_H$ is the reconstruction error whose value is 0.195.

\subsection{Modeling of ITDs}
\label{sec:itd}

\begin{figure}[!t]
 \center{\includegraphics[width=5cm] {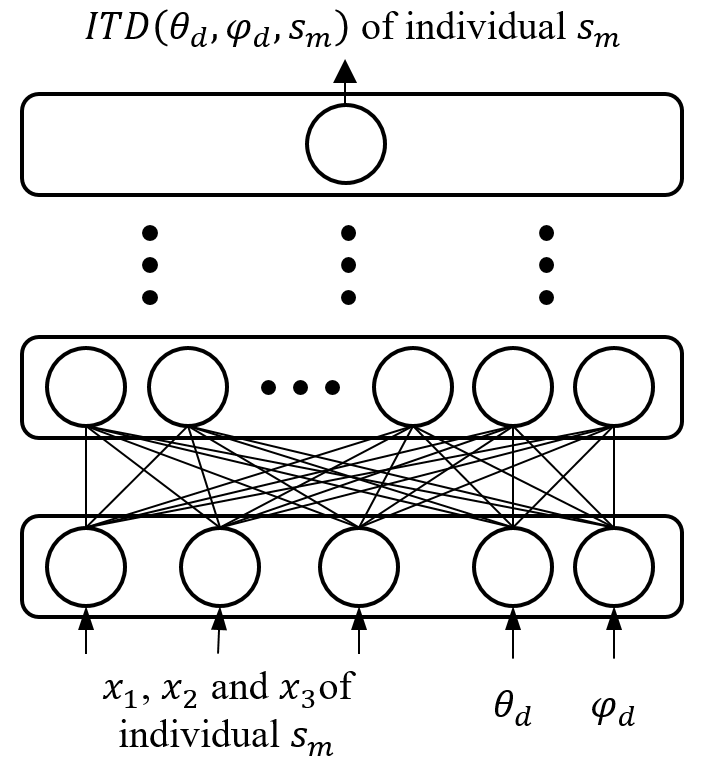}}
 \caption{\label{fig:itd-nn} The architecture of DNN for predicting the ITDs.}      
\end{figure}

As ITD not only relates to spatial directions, but also varies with different individuals, we use the anthropometric parameters and spatial directions to model it. The inputs of DNN include head width $x_1$, head height $x_2$, as well as head depth $x_3$ of an individual, target azimuth and target elevation in degrees. The individual's ITD in the target direction is taken as the ground truth, as shown in Fig. \ref{fig:itd-nn}. $x_1$, $x_2$ and $x_3$ are chosen by many experiments. The three parameters together show the best results, and adding other parameters does not yield an improvement. This indicates the strong correlation between the ITDs and the head size \cite{algazi2001estimation}. 

Similar to Section \ref{sec:spc-model}, 625 directions of ITD data for an individual are indexed and the test data are selected every four indexes. The extra directions of data are then re-indexed and we choose the validation data every five indexes. The remaining directions of data are used as training data. In the end, the training data of the 30 subjects, selected in Section \ref{sec:spca-weights}, are combined as the training set. The validation data of the 30 subjects and the test data of the remaining 7 subjects are integrated as the validation set and the test set, respectively. The mean and variance of the test set and validation set are normalized using the training set statistics to have zero mean and unit variance. Each DNN is fully connected and has three hidden layers to yield a better result. Both the activation function and the output function are set hyperbolic tangents and the learning rate is 0.001. It takes about 5500 epochs for the DNN models to converge.

\begin{figure}[!t]
 \center{\includegraphics[width=7cm] {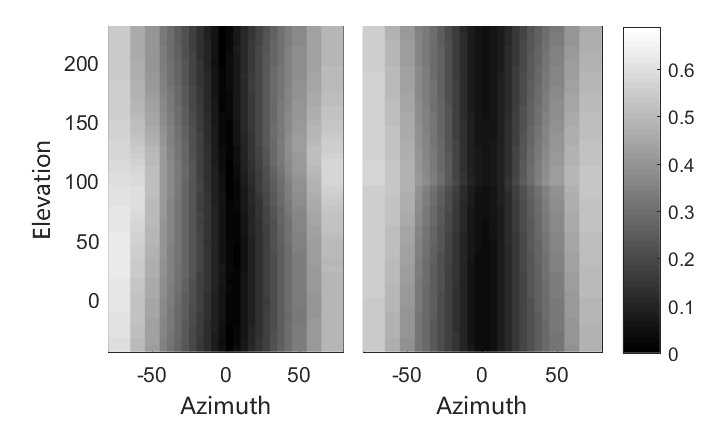}}
 \caption{\label{fig:itd} Comparison of the real value (left) and the estimated value (right) of the ITD.}      
\end{figure}

Fig. \ref{fig:itd} describes the results of ITD prediction. A total of 1250 spatial directions for a subjects' ITDs are plotted compared with its real value.

The mean absolute error (MAE) is used to evaluate the reconstruction error of ITDs. 
\begin{equation}
e_{T}=\frac{1}{S\times D}\sum\limits_{s}\sum\limits_{\theta}\sum\limits_{\phi}|\hat{T}(\theta,\phi,s_m)-T(\theta,\phi,s_m)|,
\end{equation}
where $\hat{T}$ is the estimated ITD and $T$ is the real ITD. $e_{T}$ is the reconstruction error of the ITDs, and its value is $2.22\times{10^{-2}}$ $ms$.

\subsection{Recovery of Individual HRIRs}
To reconstruct an individual's HRIRs, we first model the SPCA weights using the eight key anthropometric parameters. Then DNN is used to model SPCs, $H_\text{av}$ and ITDs respectively. The azimuth angle and elevation angle are introduced into the input layer of the DNN models, the DV-SPC, $H_\text{av}$ and ITD in arbitrary spatial directions can then be predicted.

The HRTF magnitude of a new subject $s_m$ in azimuth angle $\theta_d$ and elevation angle $\varphi_d$ can be reconstructed by solving the Eq. (\ref{equ:reconstr}). The SPCA weights for the new subject can be combined into a matrix.
\begin{equation}
\bm{d} = \left[
\begin{array}{ccc}
d_1(f_1,s_m),\quad d_2(f_1,s_m) & \dots & d_Q(f_1,s_m) \\
d_1(f_2,s_m),\quad d_2(f_2,s_m) & \dots & d_Q(f_2,s_m) \\
\vdots \qquad\quad \vdots & \vdots & \vdots \\
d_1(f_N,s_m),\quad d_2(f_N,s_m) & \dots & d_Q(f_N,s_m)
\end{array} \right],
\end{equation}
the DV-SPCs in $\theta_d$ and $\varphi_d$ is 
\begin{equation}
\bm{W}=[W_1(\theta_d,\varphi_d),W_2(\theta_d,\varphi_d),...,W_Q(\theta_d,\varphi_d)]^T, 
\end{equation}
the $H_\text{av}$ in $\theta_d$ and $\varphi_d$ is a vector of length $N$:
\begin{equation}
\bm{H_\text{AV}}=[H_\text{av}(\theta_d,\varphi_d),H_\text{av}(\theta_d,\varphi_d),...,H_\text{av}(\theta_d,\varphi_d)]^T, 
\end{equation}
then the HRTF magnitude in $\theta_d$ and $\varphi_d$ of the new subject can be restored.

The minimum phase reconstruction method is then employed to the HRTF magnitudes to generate the mono HRIRs \cite{Kistler1992A}. The ITDs obtained in Section \ref{sec:itd} are used to calculate the binaural HRIRs. The individual's HRIRs in arbitrary spatial directions can then be obtained.

\section{Objective Experiments}
\label{sec:obj}
\subsection{Evaluation using spectral distortion}
\label{subsec:obj_sd}


To evaluate the effectiveness of our proposed approach, we carried out a set of objective experiments for the proposed SPCA method, the PCA method and the generic method. The generic method used the HRTFs of the CIPIC KEMAR with small ears. The PCA method applied traditional PCA to HRTFs in each sampled spatial direction, which is described in Section \ref{sec:hrtf_data_pre}. The first twelve principal components (PCs) for all the PCA models are selected. When we apply PCA to HRTFs, for all of the spatial directions, we can restore an average of 92.02\% of the total variability for the left ear and 91.71\% for the right ear. These percentages of the total variability are close to our proposed SPCA model's selections. Then, neural network models are used to map the selected eight anthropometric parameters to the PCA weights. Thus, a total of 1250 neural network models are trained to yield the PCA weights in each spatial direction. This is quite arduous in PCA method, which will bring computation issues in the applications of HRTF prediction. Finally, individual HRTFs can be reconstructed by combining the PCs and the PCA weights.

\begin{figure}[!t]
\centering
\includegraphics[width=9cm]{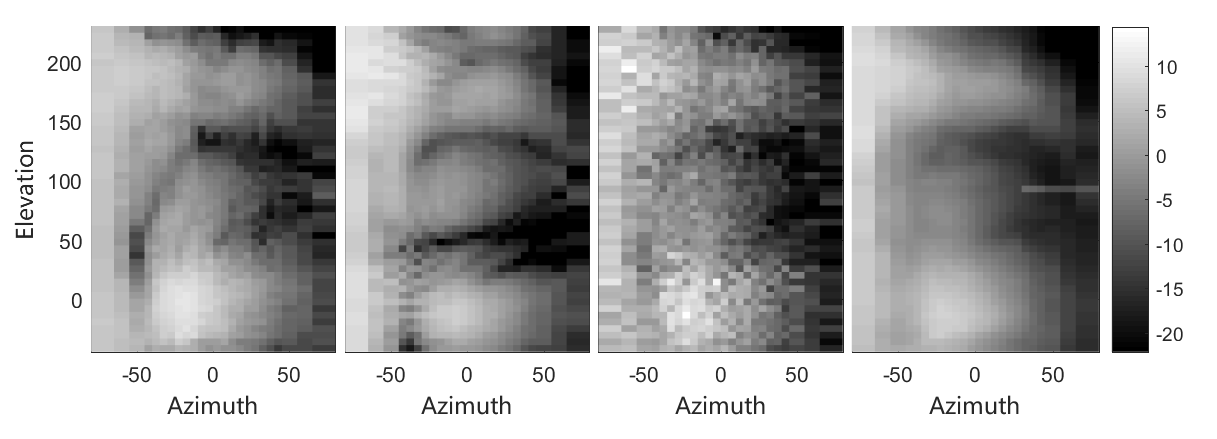}
\caption{\label{fig:SFRS} The SFRSs of the real value (1st column), the generic method(2nd column), the PCA method (3rd column) and the SPCA method (4th column) at frequency of 12.35 kHz of subject 163.} 
\end{figure}

\begin{figure}[!t]
\centering
\includegraphics[width=9cm]{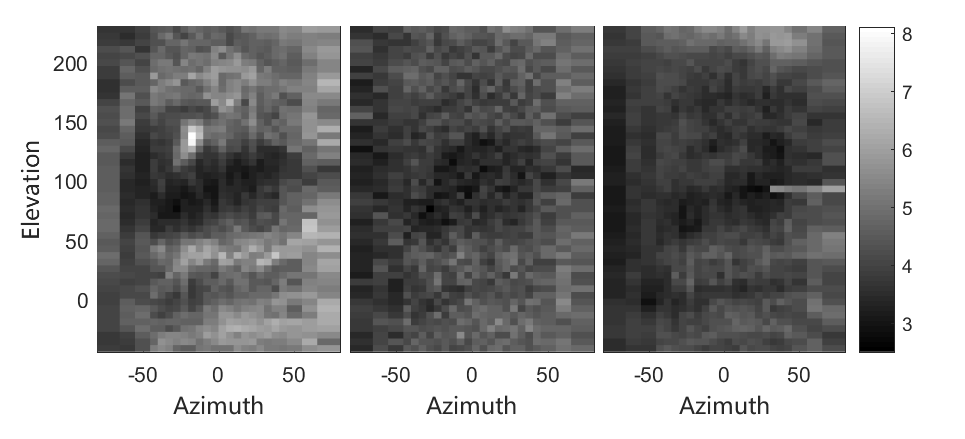}
\caption{\label{fig:SDmean7} The average prediction errors of SFRSs across all frequencies and subjects in the test set of the generic method (left), the PCA method (middle) and the SPCA method (right).} 
\end{figure}

The Spatial Frequency Response Surfaces (SFRSs) \cite{cheng1999spatial}, a spatial-domain visualization tool for HRTFs, is used to evaluate the three methods. Each frequency bin in the HRTF left or right magnitude responses constructs one SFRS, where magnitude is plotted against azimuth and elevation. Fig. \ref{fig:SFRS} shows the SFRS of the three methods compared with the real value of subject 163 at frequency of 12.35 kHz. Fig. \ref{fig:SDmean7} shows the average prediction errors of SFRS across all frequencies and subjects in the test set of the three methods. The SPCA method has smaller and more averagely distributed prediction errors than the other two methods. Namely, our proposed SPCA method has a much more smooth spatial shape than PCA method, which indicates that our algorithm predict the HRTF in spatial domain well.

Then a frequency dependent spectral distortion (SD) \cite{prepelitua2016influence} is used as an evaluation metric between the real and the test HRTF data.
\begin{equation}
SD(f, s)=\frac{1}{D}\sum\limits_{\theta}\sum\limits_{\phi}{|H(\theta, \phi, f, s)-\hat{H}(\theta, \phi, f, s)|},
\end{equation}
where $H$ is the magnitude response (dB) of the measured HRTF from the CIPIC database, and $\hat{H}$ is the magnitude response (dB) of the test HRTF.

\begin{figure}[!t]
\centering
\includegraphics[width=7cm]{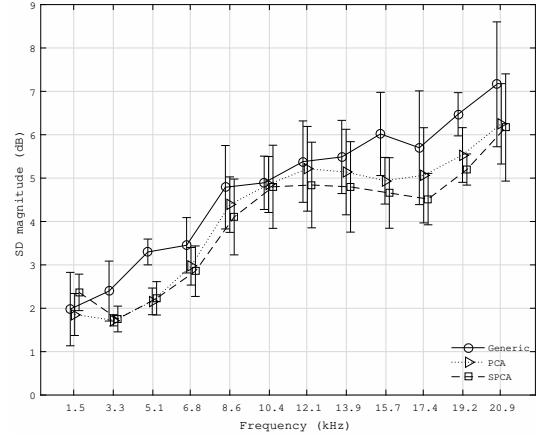}
\caption{\label{fig:SD} Comparison of the generic method, the PCA method and the SPCA method for the mean SD across all directions for some frequency bins and their deviations.} 
\end{figure}

Fig. \ref{fig:SD} shows how the SD varies with frequency of the three method. Slight translations were performed in frequency axis in order to avoid overlapping of symbols for different methods. As a result, the proposed method has the lowest SD for most frequencies compared to the other two methods. The average $SD$ of the proposed model in all the sampled directions is 5.54 dB, which is 1.13 dB lower than that of the generic method and 0.29 dB lower than that of the PCA method. T-test applied on the 12 frequency bins in Fig. \ref{fig:SD} shows that SPCA method has significant smaller SD ($p < 0.001$) than PCA method above the frequency of 6 kHz but has significant larger SD ($p < 0.001$) than PCA method blow the frequency of 6 kHz, except for 3 frequency bins, 3526 Hz ($p=0.052$), 10584 Hz ($p=0.157$) and 21168 Hz ($p=0.073$) with no significant difference. The results indicate that our proposed SPCA method predicted the HRTFs in the test set well. Since the test sets in the modeling of DV-SPCs, $H_\text{av}$ and ITDs are different from the training spatial directions, our proposed model will definitely predict the HRTFs in unmeasured spatial directions. 


\begin{figure}[!t]
\centering
\includegraphics[width=9.0cm]{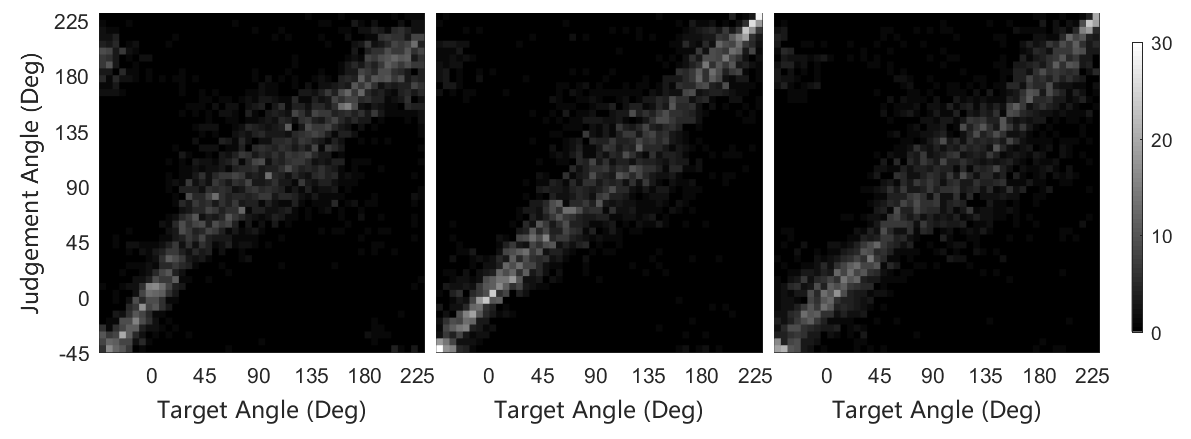}
\caption{\label{fig:aud} The localization performance of the median plane using the localization model with the generic method (left), the PCA method (middle) and the SPCA method (right).} 
\end{figure}

\subsection{Evaluation using the localization model}
\label{sec:obj-aud}

For further evaluation, an auditory based localization model \cite{zilany2014updated,majdak2014acoustic} was used to evaluate the localization performance of the generic, the PCA and SPCA calculated HRIRs. The BAUMGARTNER2014 function in the AMT tool box \cite{sondergaard2013auditory} was used in our experiment. The BAUMGARTNER2014 function is a sagittal plane localization model. As a template-based comparison model, it can be used to evaluate the performance of HRTF individualization methods. The DTFs \cite{middlebrooks1999virtual} were extracted from HRTFs \cite{majdak2014acoustic} as the inputs of the auditory model.  We apply an listener-specific sensitivity threshold of 1 and set the differential order of the spectral gradient extraction to be 0 to acquire a reasonable prediction error. Other settings of the function were set as default.

We applied the localization model on the median plane to test the performance of the generic method, the PCA method and the SPCA method. For each method, the HRIRs of all the 37 individuals used in our paper were tested. For each elevation angle, 2 runs were conducted to reduce probable prediction errors. Namely, for each elevation angle of the median plane, there were 74 predicted elevation angles of the target HRIRs in total.

\begin{table}[!t]
\caption{The average results of the objective evaluation on median plane localization experiment.}
\label{table:aud}
\centerline{
\begin{tabular}{ccccc}
\hline
\multirow{2}{*}{Method}  & \multicolumn{2}{c}{Confusion rate (\%)} & Angle of    & PE \\
            \cline{2-3}  & Up-down         & Front-back            & error (Deg) & (Deg) \\ \hline
 Generic & 11.05    & 16.32    & 17.71 $\pm$ 2.48    & 26.84 \\
 PCA     &  7.81    & 11.22    & 13.93 $\pm$ 2.21    & 21.63 \\
 SPCA    & 12.62    & 14.51    & 15.91 $\pm$ 2.53    & 24.89 \\ \hline
\end{tabular}
}
\end{table}

Fig. \ref{fig:aud} shows the localization performance of the generic method, the PCA method and the SPCA method of the median plane. Table \ref{table:aud} shows the statistical analysis of the experiment results. Four indices, including the up-down confusion rate, the front-back confusion rate, the angle of errors with standard deviations and the polar RMS (root-mean square) error (PE), were used to evaluate the localization performance. For the angle errors of the median plane, the repeated-measures ANOVA was used to verify the significance of the mean difference. According to the Shapiro-Wilk test, the data of each group obey the normal distribution ($p>0.05$). After the Mauchly's spherical hypothesis test, the variance covariance matrix of the dependent variable is not equal ($\chi^2(2)=6.279,p=0.043<0.05$). The data are corrected by Huynh-Feldt Method ($\epsilon=0.898$). The mean errors of the three methods have significant difference ($F(1.80,64.65)=34.37, p<0.001$). Bonferroni post-hoc test shows that the SPCA and PCA methods are significantly better than the generic method (both $p<0.001$). Further more, the PCA method is better than the SPCA method ($p<0.001$).

\section{Subjective Experiments}
\label{sec:sub}

In this section, the azimuth localization experiments and the elevation localization experiments were conducted to evaluate the performance of the proposed method. In the experiments, three methods were used to generate the HRTFs according to subjects' anthropometric parameters listed in Table \ref{table:parameter}. These parameters were measured before the experiments. The 4 variables listed in the 1st column were measured using a ruler with an accuracy of 2 mm. The 5 variables listed in the 3rd column were measured through photographing and manually demarcating them as shown in Fig. \ref{fig:ear}. The ruler was used to obtain the scale between the picture and real objects. All parameters were measured three times then averaged to reduce the measurement errors. The impulse responses from the headphone, used in the experiments, to the entrances of the blocked ear canals of subjects were measured using the BK Type 4101-A binaural in-ear microphone. During the experiments, headphone equalization was performed \cite{masiero2011perceptually}. 

\subsection{Azimuth Localization Experiments}
\label{sec:azi-exp}

The aim of this experiment is to compare the azimuth localization performance of the HRTFs generated by the generic method, the PCA method and the SPCA method. For the PCA method, the phases of the HRTFs are obtained by the minimum phase reconstruction method, and the ITDs are yielded using the modeling method of Section \ref{sec:itd}.

The stimulus in this experiment was a train of eight 250-ms bursts of Gaussian noise (20-ms cosine-squared onset-offset ramps), with 300 ms of silence between the bursts. The HRIRs of twelve azimuth angles, 0, 30, 55, 80, 125, 150, 180, 210, 235, 280, 305 and 330 degrees, in three elevation angles, 0, 22.5 and 45 degrees, are generated by the generic method, the PCA method and the SPCA method respectively. Note that the HRIRs of the four azimuth angles, 55, 150, 210 and 305 degrees, in the three elevation angles are not in the training set and the validation set of each DNN model, i.e. Section \ref{sec:spc-model}, Section \ref{sec:msf-model} and Section \ref{sec:itd}. Then, the stimulus is filtered by the HRIRs produced by the three methods to create three kinds of sounds.

\begin{figure}[!t]
\begin{minipage}[b]{0.5\linewidth}
\centering
\includegraphics[width=1.7in]{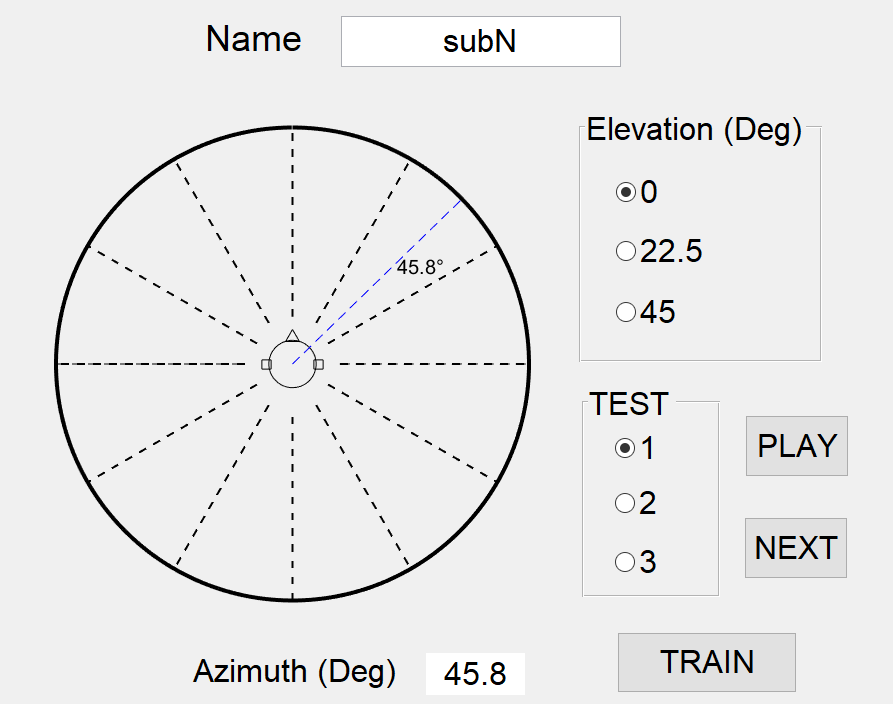}
\subfigure{\scriptsize{(a) Interface for azimuth experiments}}
\end{minipage}%
\begin{minipage}[b]{0.5\linewidth}
\centering
\includegraphics[width=1.61in]{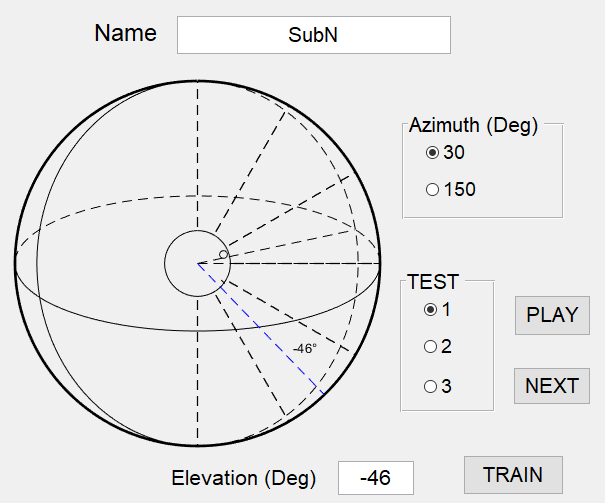}
\subfigure{\scriptsize{(b) Interface for elevation experiments}}
\end{minipage}
\caption{\label{fig:interface} The user interfaces for subjects to give percepted directions of sounds.}
\end{figure}

A total of three azimuth localization experiments are performed, and each experiment includes three tests. The three experiments correspond to three elevation angles, 0, 22.5 and 45 degrees, respectively. Each test is a kind of sound generated by one method, and the order of the three tests is arranged by latin square design across every three subjects. Before each test, the subject is trained using the test sound of the other eight azimuth angles, 0, 45, 90, 135, 180, 225, 270 and 315 degrees. Through listening to these sounds, the subject can build up the spatial perception for this kind of virtual sound. After that, thirty-six binaural sounds are randomly played to the subject by a Sennheiser HD 650 headphone through a Sound Blaster sound card. The thirty-six sounds contain twelve directions' sounds, and each direction appears three times. The subject can listen to one sound many times until he/she can identify the exact perceived direction. The subject gave the exact direction of each sound he/she perceived during the test through an interface on a computer, which is shown in Fig. \ref{fig:interface} (a). The blue dashed line in the interface is moved with the mouse cursor and indicates the exact angle. After each experiment, there were five minutes for a break. 

Eighteen subjects (11 male 7 female, age from 21 to 27) with normal hearing took part in the experiments. All of the experiments were performed in a sound booth (Background noise level: 20.9 dBA), with no light during each test.

\begin{figure}[!t]
 \center{\includegraphics[width=9.0cm] {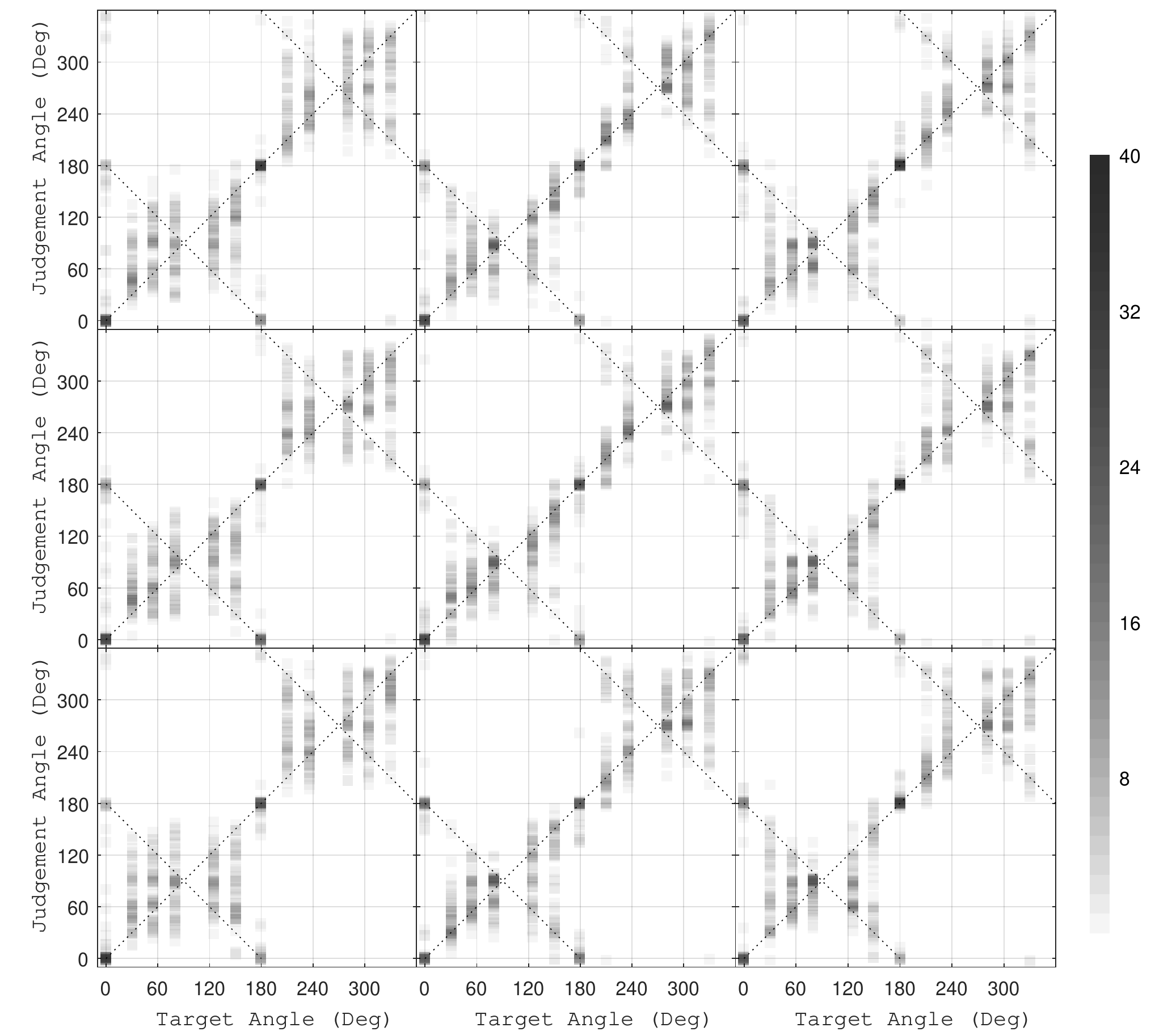}}
 \caption{\label{fig:azi1} Judged direction versus target direction of all subjects using the generic method (left column), the PCA method (middle column) and the SPCA method (right column) in elevations of 0 degrees (top row), 22.5 degrees (middle row) and 45 degrees (bottom row). Two oblique lines with a slope of 135 degrees correspond to the front-back confusions.}      
\end{figure}

Fig. \ref{fig:azi1} shows the results of the localization experiments of all eighteen subjects in three elevation angles respectively. The judgments are plotted as a function of the coordinates of the targets. The left column, the middle column and the right column depict the judgments using the generic method, the PCA method and the SPCA method respectively. There are 648 judgments shown in each panel, corresponding to the 54 judgments made for each of the twelve binaural sounds.

Note that the localization performance of the directions in the test set of DNN modeling is as good as the directions in the training set. The localization performances of the PCA method and the SPCA method are better than that of the generic method. More judgments fall upon the diagonal line in the PCA and the SPCA method, which indicates a higher precision of localization. The reconstruction error of ITD is larger than the just noticeable differences (JNDs) for some specific directions. For example, a few subjects perceived that azimuth of 0 degrees is not directly ahead.

The front-back confusion rate and error of angle with their standard deviations of the azimuth localization experiments are shown in Table \ref{table:azi}. Angle of error is the angle difference between the target angle and the judgement angle. Front-back confusion is corrected before calculating the difference. The repeated-measures ANOVA was used to compare the localization performance (confusion rate and error of angle) of the three methods (generic, PCA and SPCA) for three elevations (0, 22.5 and 45 degrees). According to the Shapiro-Wilk test, all groups of data obey the normal distribution ($P>0.05$). Through the Mauchly's spherical hypothesis test, the variance covariance matrix of the dependent variable is dominantly equal ($p>0.05$ except for the condition of error of angle in 22.5 degrees, $\chi^2(2)=7.97,p<0.05$). The data, which doesn't satisfy spherical assumption, are corrected by Greenhouse-geisser Method ($\epsilon=0.718$). For confusion rate, there is no significant difference for three methods in elevation of 0 and 45 degrees ($F(2,34)=2.41, p=0.11$ and $F(2,34)=2.10, p=0.14$), but a significant difference in elevation of 22.5 degrees ($F(2,34)=8.71, p<0.005$). Bonferroni post-hoc test for elevation of 22.5 degrees shows that the PCA method has significantly smaller confusion rate than the SPCA and generic methods ($p<0.05$ and $p<0.005$);  the confusion rates of the SPCA method and the generic method have no significant difference ($p=0.92$). For error of angle, there are significant differences for all three elevations ($F(2,34)=13.23, p<0.001$, $F(1.436,24.42)=17.38, p<0.001$ and $F(2,34)=16.25, p<0.001$): the generic method has significantly larger errors than the PCA and SPCA methods ($p<0.005$ and $p<0.005$ for all three elevations); the PCA method and SPCA method have no significant difference ($p=1.0$ for all three elevations). 

\begin{table}[!t]
\caption{The average results of the azimuth localization experiments.}
\label{table:azi}
\centerline{
\begin{tabular}{cccc}
\hline
\begin{tabular}[c]{@{}c@{}}Elevation\\ angle (Deg)\end{tabular}     & Method  & \begin{tabular}[c]{@{}c@{}}Front-back\\ confusion rate (\%)\end{tabular} & \begin{tabular}[c]{@{}c@{}}Angle of\\ error (Deg)\end{tabular} \\ \hline
\multirow{3}{*}{   0}   & Generic & 28.2 $\pm$ 9.6    & 18.90 $\pm$ 4.25 \\
                        & PCA     & 22.4 $\pm$ 11.1    & 14.84 $\pm$ 3.63 \\
                        & SPCA    & 24.5 $\pm$ 12.7    & 15.15 $\pm$ 4.04 \\ \hline
\multirow{3}{*}{22.5}   & Generic & 29.0 $\pm$ 12.2    & 19.49 $\pm$ 4.22 \\
                        & PCA     & 17.0 $\pm$ 9.9    & 15.25 $\pm$ 3.25 \\
                        & SPCA    & 25.6 $\pm$ 13.3    & 14.96 $\pm$ 3.74 \\ \hline
\multirow{3}{*}{  45}   & Generic & 31.8 $\pm$ 11.9    & 19.79 $\pm$ 3.30 \\
                        & PCA     & 25.3 $\pm$ 10.8    & 15.74 $\pm$ 3.81 \\
                        & SPCA    & 29.0 $\pm$ 10.7    & 15.28 $\pm$ 3.75 \\ \hline
\end{tabular}
}
\end{table}

\subsection{Elevation Localization Experiments}
\label{sec:sub-elev}
The purpose of this experiment is to compare the elevation localization performance of the HRIRs generated by the generic method, the PCA method and the SPCA method. 

\begin{figure}[!t]
\begin{minipage}[b]{0.5\linewidth}
\centering
\includegraphics[width=4.5cm]{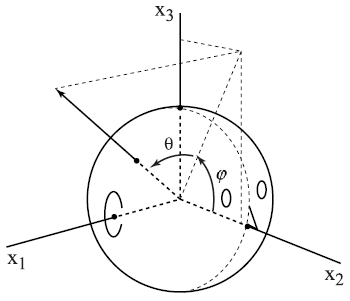}
\subfigure{\scriptsize{(a)}}
\end{minipage}%
\begin{minipage}[b]{0.5\linewidth}
\centering
\includegraphics[width=4.5cm]{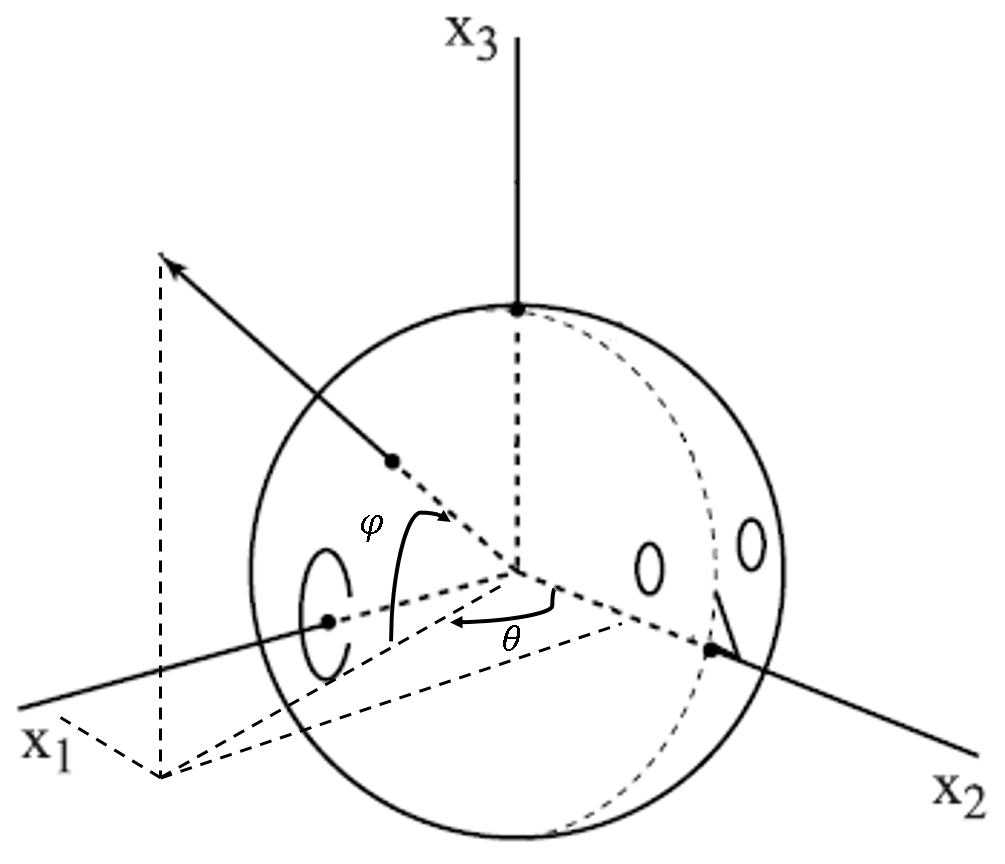}
\subfigure{\scriptsize{(b)}}
\end{minipage}
\caption{\label{fig:aural} Definition of (a) interaural-polar coordinate system and (b) head-related spherical coordinate system.}
\end{figure}

In the CIPIC database, the azimuth angle and the elevation angle are measured in a head-centered interaural-polar coordinate system shown in Fig. \ref{fig:aural} (a). However, the elevation angle in a spherical coordinate system is much closer to our auditory perception. Therefore, the azimuth angle and the elevation angle in the CIPIC database are represented in the head-related spherical coordinate system plotted in Fig. \ref{fig:aural} (b). The transformation formulas are as follows: 
\begin{equation}
\begin{split}
&\sin(\theta')=\sin(\theta)\cos(\varphi),\\
&\tan(\varphi')={\tan(\varphi)} \left/ {\cos(\theta)}\right.,
\end{split}
\end{equation}
where $\theta$ and $\varphi$ refer to the azimuth angle and the elevation angle in the head-related spherical coordinate system respectively, and $\theta'$ and $\varphi'$ are the azimuth angle and the elevation angle in the interaural-polar coordinate system respectively.

In the elevation localization experiments, we use the same stimulus in the Section \ref{sec:azi-exp}. The HRIRs of seven elevation angles, 90, 70, 47, 19, 0, -19 and -42 degrees, in two azimuth angles, nearly 30 and 150 degrees, are generated by the generic method, the PCA method and the SPCA method respectively. Then, the stimulus is filtered by the HRIRs produced by the three methods to create three kinds of sounds.

A total of two elevation localization experiments are performed, and each experiment includes three tests. The two experiments correspond to two azimuth angles, 30 and 150 degrees, respectively. The order of the two experiments is balanced. Each test is a kind of sound generated by one method, and the order of the three tests is arranged by latin square design across every three subjects. Before each test, the subject is trained using the test sound of five elevation angles, 90, 58, 30, 0 and -30 degrees. Through listening to these sounds, the subject gradually builds up the spatial perception for this kind of virtual sound. After that, twenty-one binaural sounds are randomly played to the subject by a Sennheiser HD 650 headphone through a Sound Blaster sound card. The twenty-one sounds contain seven directions' sounds, and each direction appears three times. The subject can listen to one sound many times until he/she can identify the exact direction. The subject gave the exact direction of each sound he/she perceived during the test through an interface on a computer, which is shown in Fig. \ref{fig:interface} (b). The blue dashed line in the interface is moved with the mouse cursor and indicates the exact angle. After each experiment, there were five minutes for a break. 

\begin{figure}[!t]
 \center{\includegraphics[width=9.0cm] {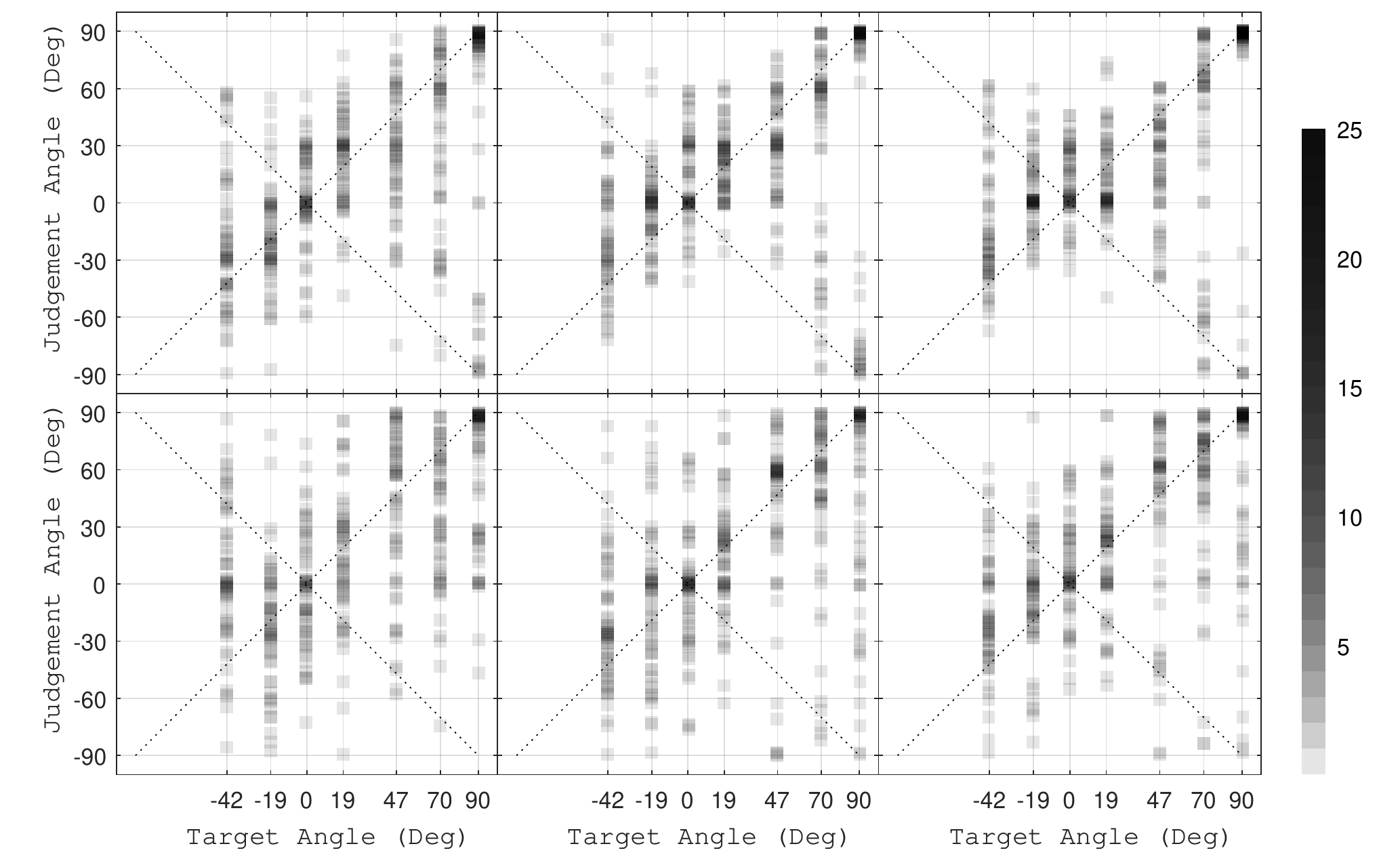}}
 \caption{\label{fig:elev1} Judged direction versus target direction of all subjects using the generic method (left column), the PCA method (middle column) and the SPCA method (right column) in azimuth of 30 degrees (top row) and 150 degrees (bottom row). The oblique line with a slope of 135 degrees corresponds to the up-down confusions.}
\end{figure}

Eighteen subjects (the same as in the azimuth localization experiments) with normal hearing took part in the experiments. All experiments were performed in a sound booth (Background noise level: 20.9 dBA), with no light during each test.

Fig. \ref{fig:elev1} shows the results of the localization experiments of all eighteen subjects in two azimuth angles respectively. The judgments are plotted as a function of the coordinates of the targets. The left column, the middle column and the right column depict the judgments using the generic method, the PCA method and the SPCA method respectively. There are 378 judgments shown in each panel, corresponding to the 54 judgments made to each of the seven binaural sounds.


\begin{table}[!t]
\caption{The average results of the elevation localization experiments.}
\label{table:elev}
\centerline{
\begin{tabular}{ccccc}
\hline
\begin{tabular}[c]{@{}c@{}}Azimuth\\ angle (Deg)\end{tabular}  & Method  & \begin{tabular}[c]{@{}c@{}}Up-down\\ confusion rate (\%)\end{tabular} & \begin{tabular}[c]{@{}c@{}}Angle of\\ error (Deg)\end{tabular} \\ \hline
\multirow{3}{*}{ 30}    & Generic & 16.1 $\pm$ 12.1    & 18.69 $\pm$ 6.42 \\
                        & PCA     & 18.8 $\pm$ 14.2    & 16.65 $\pm$ 3.75 \\
                        & SPCA    & 20.4 $\pm$ 10.3    & 15.04 $\pm$ 4.13 \\ \hline
\multirow{3}{*}{150}    & Generic & 17.2 $\pm$ 9.3    & 25.00 $\pm$ 8.96 \\
                        & PCA     & 18.0 $\pm$ 13.0    & 21.86 $\pm$ 7.86 \\
                        & SPCA    & 16.7 $\pm$ 14.8    & 20.15 $\pm$ 9.02 \\ \hline
\end{tabular}
}
\end{table}


The up-down confusion rate and error of angle with their standard deviations of the elevation localization experiments are shown in Table \ref{table:elev}. Angle of error is the angle difference between the target angle and the judgement angle. Up-down confusion is corrected before calculating the difference. The repeated-measures ANOVA was used to compare the localization performance (confusion rate and error of angle) of the three methods (generic, PCA and SPCA) for two azimuths (30 and 150 degrees). According to the Shapiro-Wilk test, all groups of data obey the normal distribution ($p>0.05$). Through the Mauchly's spherical hypothesis test, the variance covariance matrix of the dependent variable is dominantly equal ($p>0.05$ except for the condition of error of angle in 30 degrees, $\chi^2(2)=9.11,p<0.05$). The data, which doesn't satisfy spherical assumption, are corrected by Greenhouse-geisser Method ($\epsilon=0.697$). For confusion rates, there is no significant difference for three methods in two azimuths ($F(2,34)=0.70, p=0.51$ and $F(2,34)=0.08, p=0.92$). For error of angle, there are significant differences for three methods in two azimuths ($F(1.39,23.7)=3.92, p<0.05$ and $F(2,34)=4.76, p<0.05$). With Bonferroni post-hoc test, the generic method has significantly larger errors than the SPCA method ($p<0.05$ for two azimuths); there is no significant difference between the PCA method and the generic and SPCA methods for two azimuths (PCA VS generic: $p=0.73$ and $p=0.15$; PCA VS SPCA: $p=0.45$ and $p=1.0$).


Based on both azimuth and elevation localization experiments, we conclude that both the SPCA and PCA methods are superior than the generic method. For most of the localization experiments, the SPCA method and the PCA method have similar performances, except for the confusion rate in the elevation of 22.5 degrees, where PCA method has significantly smaller confusion rate than the SPCA method. However, the proposed SPCA method predicts the HRTFs outside the training data well, which indicates that the SPCA method can predict the HRTFs in unmeasured spatial directions.

\section{Conclusion}
\label{sec:conclusion}

This paper proposed the individual HRTFs modeling method using DNN based on SPCA. By modeling the SPCs, the SPCA weights, the $H_\text{av}$ and the ITDs respectively, we reconstruct the individual HRIRs in arbitrary spatial directions. The objective and subjective experiments are performed to evaluate the individual HRTFs generated by the proposed method, the PCA method, and the generic method. Both the objective and subjective experiments\rq{} results show that the proposed SPCA method and PCA method are superior than the generic method. The spectral distortion of the SPCA method is significantly smaller than PCA method in high frequencies but significantly larger in low frequencies, except for few frequencies with no significant difference. The evaluation using the localization model shows that the PCA method is better than the SPCA method. The subjective experiments show the SPCA method and the PCA method have similar performances, except for the confusion rate in the elevation of 22.5 degrees, where PCA method has a better performance than the SPCA method. Nevertheless, the results indicate that our proposed SPCA method could predict the HRTFs in arbitrary spatial directions well. For the future work, we will use more HRTF data to implement the individual HRTF models for better performance.

\ifCLASSOPTIONcaptionsoff
  \newpage
\fi



%



\bibliographystyle{IEEEtran}
\bibliography{IEEEabrv,refs}

\begin{thebibliography}{10}
\providecommand{\url}[1]{#1}
\csname url@samestyle\endcsname
\providecommand{\newblock}{\relax}
\providecommand{\bibinfo}[2]{#2}
\providecommand{\BIBentrySTDinterwordspacing}{\spaceskip=0pt\relax}
\providecommand{\BIBentryALTinterwordstretchfactor}{4}
\providecommand{\BIBentryALTinterwordspacing}{\spaceskip=\fontdimen2\font plus
\BIBentryALTinterwordstretchfactor\fontdimen3\font minus
  \fontdimen4\font\relax}
\providecommand{\BIBforeignlanguage}[2]{{%
\expandafter\ifx\csname l@#1\endcsname\relax
\typeout{** WARNING: IEEEtran.bst: No hyphenation pattern has been}%
\typeout{** loaded for the language `#1'. Using the pattern for}%
\typeout{** the default language instead.}%
\else
\language=\csname l@#1\endcsname
\fi
#2}}
\providecommand{\BIBdecl}{\relax}
\BIBdecl

\bibitem{blauert1997spatial}
J.~Blauert, \emph{Spatial hearing: the psychophysics of human sound
  localization}.\hskip 1em plus 0.5em minus 0.4em\relax MIT press, 1997.

\bibitem{Wenzel1991Localization}
E.~Wenzel, F.~Wightman, and D.~Kistler, ``Localization with non-individualized
  virtual acoustic display cues,'' in \emph{Sigchi Conference on Human Factors
  in Computing Systems}, 1991, pp. 351--359.

\bibitem{wenzel1993localization}
E.~M. Wenzel, M.~Arruda, D.~J. Kistler, and F.~L. Wightman, ``Localization
  using nonindividualized head-related transfer functions,'' \emph{The Journal
  of the Acoustical Society of America}, vol.~94, no.~1, pp. 111--123, 1993.

\bibitem{algazi2001cipic}
V.~R. Algazi, R.~O. Duda, D.~M. Thompson, and C.~Avendano, ``The cipic hrtf
  database,'' in \emph{2001 IEEE Workshop on the Applications of Signal
  Processing to Audio and Acoustics}.\hskip 1em plus 0.5em minus 0.4em\relax
  New Platz, NY, USA,: IEEE, 2001, pp. 99--102.

\bibitem{Qu2009Distance}
T.~Qu, Z.~Xiao, M.~Gong, Y.~Huang, X.~Li, and X.~Wu, ``Distance-dependent
  head-related transfer functions measured with high spatial resolution using a
  spark gap,'' \emph{IEEE Transactions on Audio, Speech, and Language
  Processing}, vol.~17, no.~6, pp. 1124--1132, 2009.

\bibitem{Kreuzer2009Fast}
W.~Kreuzer, P.~Majdak, and Z.~Chen, ``Fast multipole boundary element method to
  calculate head-related transfer functions for a wide frequency range,''
  \emph{Journal of the Acoustical Society of America}, vol. 126, no.~3, pp.
  1280--90, 2009.

\bibitem{ma2015finite}
F.~Ma, J.~Wu, M.~Huang, W.~Zhang, W.~Hou, and C.~Bai, ``Finite element
  determination of the head-related transfer function,'' \emph{Journal of
  Mechanics in Medicine and Biology}, vol.~15, no.~05, p. 1550066, 2015.

\bibitem{Xiao2003Finite}
T.~Xiao and Q.~Liu, ``Finite difference computation of head-related transfer
  function for human hearing.'' \emph{Journal of the Acoustical Society of
  America}, vol. 113, no.~5, pp. 2434--41, 2003.

\bibitem{brown1998structural}
C.~P. Brown and R.~O. Duda, ``A structural model for binaural sound
  synthesis,'' \emph{IEEE transactions on speech and audio processing}, vol.~6,
  no.~5, pp. 476--488, 1998.

\bibitem{Middlebrooks1999Individual}
J.~C. Middlebrooks, ``Individual differences in external-ear transfer functions
  reduced by scaling in frequency.'' \emph{Journal of the Acoustical Society of
  America}, vol. 106, no.~3, pp. 1480--1492, 1999.

\bibitem{Zotkin2003HRTF}
D.~N. Zotkin, J.~Hwang, R.~Duraiswaini, and L.~S. Davis, ``Hrtf personalization
  using anthropometric measurements,'' in \emph{2003 IEEE Workshop on
  Applications of Signal Processing To Audio and Acoustics}, New Paltz, NY,
  USA, 2003, pp. 157--160.

\bibitem{jin2000enabling}
C.~Jin, P.~Leong, J.~Leung, A.~Corderoy, and S.~Carlile, ``Enabling
  individualized virtual auditory space using morphological measurements,'' in
  \emph{Proceedings of the First IEEE Pacific-Rim Conference on Multimedia},
  Sydney, Australia, 2000, pp. 235--238.

\bibitem{Hu2008HRTF}
H.~Hu, L.~Zhou, H.~Ma, and Z.~Wu, ``Hrtf personalization based on artificial
  neural network in individual virtual auditory space,'' \emph{Applied
  Acoustics}, vol.~69, no.~2, pp. 163--172, 2008.

\bibitem{guezenoc2018hrtf}
C.~Guezenoc and R.~Seguier, ``Hrtf individualization: A survey,'' in
  \emph{Audio Engineering Society Convention 145}.\hskip 1em plus 0.5em minus
  0.4em\relax Audio Engineering Society, 2018.

\bibitem{Fink2012Tuning}
K.~J. Fink and L.~Ray, ``Tuning principal component weights to individualize
  hrtfs,'' in \emph{IEEE International Conference on Acoustics, Speech and
  Signal Processing}, Kyoto, Japan, 2012, pp. 389--392.

\bibitem{luo2013virtual}
Y.~Luo, D.~N. Zotkin, and R.~Duraiswami, ``Virtual autoencoder based
  recommendation system for individualizing head-related transfer functions,''
  in \emph{IEEE Workshop on Applications of Signal Processing to Audio and
  Acoustics}, New York, USA, 2013, pp. 1--4.

\bibitem{Xie2012Recovery}
B.~Xie, ``Recovery of individual head-related transfer functions from a small
  set of measurements.'' \emph{Journal of the Acoustical Society of America},
  vol. 132, no.~1, pp. 282--294, 2012.

\bibitem{romigh2015efficient}
G.~D. Romigh, D.~S. Brungart, R.~M. Stern, and B.~D. Simpson, ``Efficient real
  spherical harmonic representation of head-related transfer functions,''
  \emph{IEEE Journal of Selected Topics in Signal Processing}, vol.~9, no.~5,
  pp. 921--930, 2015.

\bibitem{smith1983techniques}
J.~O. Smith, ``Techniques for digital filtering design and system
  identification with the violin,'' Ph.D. dissertation, CCRMA, Stanford, 1983.

\bibitem{jolliffe2003principal}
I.~Jolliffe, ``Principal component analysis,'' \emph{Technometrics}, vol.~45,
  no.~3, p. 276, 2003.

\bibitem{Kistler1992A}
D.~J. Kistler and F.~L. Wightman, ``A model of head-related transfer functions
  based on principal components analysis and minimum-phase reconstruction,''
  \emph{Journal of the Acoustical Society of America}, vol.~91, no.~3, p. 1637,
  1992.

\bibitem{Zhang2011Statistical}
M.~Zhang, R.~A. Kennedy, T.~D. Abhayapala, and W.~Zhang, ``Statistical method
  to identify key anthropometric parameters in hrtf individualization,'' in
  \emph{The Workshop on Hands-Free Speech Communication \& Microphone Arrays},
  Edinburgh, UK, 2011, pp. 213--218.

\bibitem{Zeng2010A}
X.~Zeng, S.~Wang, and L.~Gao, ``A hybrid algorithm for selecting head-related
  transfer function based on similarity of anthropometric structures,''
  \emph{Journal of Sound \& Vibration}, vol. 329, no.~19, pp. 4093--4106, 2010.

\bibitem{Nicol2006Looking}
R.~Nicol, V.~Lemaire, A.~Bondu, and S.~Busson, ``Looking for a relevant
  similarity criterion for hrtf clustering: a comparative study,'' in
  \emph{Audio Engineering Society Convention 120}.\hskip 1em plus 0.5em minus
  0.4em\relax Paris, France: Audio Engineering Society, 2006, p. 6653.

\bibitem{IMM2012-06284}
R.~B. Palm, ``Prediction as a candidate for learning deep hierarchical models
  of data,'' Master's thesis, 2012.

\bibitem{algazi2001estimation}
V.~R. Algazi, C.~Avendano, and R.~O. Duda, ``Estimation of a spherical-head
  model from anthropometry,'' \emph{Journal of the Audio Engineering Society},
  vol.~49, no.~6, pp. 472--479, 2001.

\bibitem{cheng1999spatial}
C.~I. Cheng and G.~H. Wakefield, ``Spatial frequency response surfaces (sfr's):
  An alternative visualization and interpolation technique for head-related
  transfer functions (hrtf's),'' in \emph{Audio Engineering Society Conference:
  16th International Conference: Spatial Sound Reproduction}.\hskip 1em plus
  0.5em minus 0.4em\relax Audio Engineering Society, 1999.

\bibitem{prepelitua2016influence}
S.~Prepeliț{\u{a}}, M.~Geronazzo, F.~Avanzini, and L.~Savioja, ``Influence of
  voxelization on finite difference time domain simulations of head-related
  transfer functions,'' \emph{The Journal of the Acoustical Society of
  America}, vol. 139, no.~5, pp. 2489--2504, 2016.

\bibitem{zilany2014updated}
M.~S.~A. Zilany, I.~C. Bruce, and L.~H. Carney, ``Updated parameters and
  expanded simulation options for a model of the auditory periphery,''
  \emph{The Journal of the Acoustical Society of America}, vol. 135, no.~1, pp.
  283--286, 2014.

\bibitem{majdak2014acoustic}
P.~Majdak, R.~Baumgartner, and B.~Laback, ``Acoustic and non-acoustic factors
  in modeling listener-specific performance of sagittal-plane sound
  localization,'' \emph{Frontiers in psychology}, vol.~5, p. 319, 2014.

\bibitem{sondergaard2013auditory}
P.~L. S{\o}ndergaard and P.~Majdak, ``The auditory modeling toolbox,'' in
  \emph{The technology of binaural listening}.\hskip 1em plus 0.5em minus
  0.4em\relax Springer, 2013, pp. 33--56.

\bibitem{middlebrooks1999virtual}
J.~C. Middlebrooks, ``Virtual localization improved by scaling
  nonindividualized external-ear transfer functions in frequency,'' \emph{The
  Journal of the Acoustical Society of America}, vol. 106, no.~3, pp.
  1493--1510, 1999.

\bibitem{masiero2011perceptually}
B.~Masiero and J.~Fels, ``Perceptually robust headphone equalization for
  binaural reproduction,'' in \emph{Audio Engineering Society Convention
  130}.\hskip 1em plus 0.5em minus 0.4em\relax Audio Engineering Society, 2011.

\end{thebibliography}

%






\end{document}